\documentclass[a4paper,twocolumn,11pt]{quantumarticle}
\pdfoutput=1
\usepackage[utf8]{inputenc}
\usepackage[english]{babel}
\usepackage[T1]{fontenc}
\usepackage{amsmath}
\usepackage{hyperref}

\usepackage{tikz}
\usepackage{lipsum}

\usepackage{xcolor}
\newcommand{\bra}[1]{\langle #1|}
\newcommand{\ket}[1]{|#1\rangle}

\begin{document}

\title{Quantum digital simulations of time-domain semiconductor spectroscopy in the semi-classical approximation} 

\author{Mykhailo Klymenko}
\email{Contact author: mike.klymenko@monash.edu}
\altaffiliation{Department of Materials Science and Engineering, Monash University, 20 Research Way, New Horizons, Clayton 3800, Victoria, Australia}
\author{Bahar Goldozian}%
\affiliation{CSIRO, Research Way, Clayton 3168, Victoria, Australia}
\author{Thong Hoang}%
\affiliation{CSIRO, Research Way, Clayton 3168, Victoria, Australia}
\author{Jared H. Cole}%
\affiliation{Department of Physics, School of Science, RMIT University, Melbourne, Victoria, Australia}
\author{Muhammad Usman}%
\affiliation{CSIRO, Research Way, Clayton 3168, Victoria, Australia}
\affiliation{School of Physics, The University of Melbourne, Parkville 3010, Victoria, Australia}

\begin{abstract}
We present a digital quantum simulation framework for ultrafast optical spectroscopy of semiconductor materials. The framework is based on Brillouin-zone discretization and is designed as a quantum alternative to classical simulations based on the semiconductor Bloch equations. Its current capabilities include quantum simulations of linear absorption and optical gain, incorporating Lorentzian broadening, finite-temperature band-filling effects, and reduced-dimensionality effects. Comparison with classical simulations of GaAs within the single-particle approximation demonstrates quantitative agreement in the noiseless limit. The inclusion of realistic hardware noise characteristic of NISQ-era quantum computers effectively manifests as an additional source of scattering, resulting in increased spectral broadening. We also show that the linear optical response, including many-body interactions treated within the Hartree–Fock approximation, can be mapped onto a system of coupled classical oscillators. Under certain assumptions about the query complexity of the oscillator couplings, this system exhibits a provable quantum advantage, as demonstrated in [Phys. Rev. X 13, 041041 (2023)]. Beyond the Hartree–Fock approximation, classical simulations based on many-body quantum kinetic equations face the so-called hierarchy problem, whereas the algorithmic complexity of the quantum simulations remains unchanged, offering a promising approach to overcoming this challenge.
\end{abstract}

\maketitle


\section{\label{sec:intro} Introduction}

Semiconductor optical spectroscopy employs optical characterization techniques (e.g. absorption and photoluminescence spectroscopy, including linear and nonlinear cases, as well as continuous-wave and ultrafast implementations) to probe the fundamental properties of semiconductor materials -- including electronic structure, deformation effects, chemical composition, carrier dynamics and transport -- ranging from bulk crystals to nanostructures \cite{jimenez2018spectroscopic, kalt2019semiconductor, chow1999semiconductor, haug_koch}. These measurements are essential for both research and industrial quality control, enabling the development of advanced semiconductor electronic, optoelectronic and photonic devices \cite{piprek2005optoelectronic, Nisoli_2022}. 

Simulating spectroscopy experiments plays a crucial role in understanding and accurately interpreting the experimentally obtained spectral characteristics. The physical principles of semiconductor spectroscopy are based on light-matter interactions and involve non-equilibrium dynamics of quantum fields with many-body interactions (electron-electron correlations, scatterings on phonons, dynamical dielectric screening, etc.). Classical simulations of such systems typically involve solving many-body quantum kinetic equations that describe the non-equilibrium dynamics of the system and are based either on non-equilibrium Green’s functions \cite{kadanoff1962quantum,Stefanucci2013,bonitz2016quantum} or the semiconductor Bloch equations \cite{haug_koch, chow1999semiconductor, KIRA2006155}. In both cases, this results in a large system of strongly coupled integro-differential equations. In the case of many-body interactions treated without approximations, the number of equations becomes infinite, which is known as the hierarchy problem \cite{KIRA2006155}. The simulation of such dynamics is, in general, intractable for classical computers and therefore requires approximating approaches. However, even with approximations, obtaining practically useful results still demands enormous computational resources and access to high-performance computing infrastructure. Motivated by recent advances in quantum computing hardware, we believe that this problem can be efficiently solved using quantum computers.

The idea of simulating quantum systems using quantum computers originates from Feynman’s 1982 talk  \cite{Feynman1982}, in which he considered the possibility and advantages of simulating quantum systems using another quantum system - a quantum computing machine. In his talk, Feynman proposed that, unlike classical computers, a quantum computer can simulate another quantum system exactly. Later, Seth Lloyd has formally demonstrated quantum advantage for digital quantum simulation \cite{Lloyd1996}.  Beyond the digital quantum computing, a significant breakthrough in quantum simulation was the development of analog simulators based on ultracold atoms in optical lattices, which have been applied to strongly correlated systems \cite{Dalmonte2018,Arguello-Luengo2019,Daley2022,3nx4-bnyy}. 

Nevertheless, most state-of-the-art results in applications of quantum computing in physics and chemistry are based on variational algorithms and focus on stationary ground and excited states \cite{bauer2020quantum, McArdle2020, Sherbert2022, Ma2020}. Only recently has attention begun to shift toward simulations of time evolution \cite{PRXQuantum.5.040316, Zylberman_2026} and finite-temperature systems, for which Gibbs-state sampling algorithms have been proposed \cite{chowdhury2020variational, rajakumar2026gibbs,Chen2025}. In particular, quantum digital simulations have been applied for tunneling spectroscopy of finite spin chains \cite{PhysRevResearch.4.043106}, nuclear magnetic resonance spectroscopy \cite{fratus2026}, and molecular two-dimensional electronic spectroscopy \cite{Bruschi2024}. Among its key contributions, Ref.~\cite{PhysRevResearch.4.043106} demonstrates that efficient spectroscopy simulation constitutes a powerful paradigm for addressing eigenvalue problems in quantum many-body physics and quantum chemistry. Indeed, whether physically implemented or simulated, spectroscopy can provide the same information about ground- and excited-state energies as computational approaches in both classical and quantum computing. We anticipate that in the post-NISQ era, simulation-based approaches will increasingly supplant variational methods, as they require fewer encoding and measurement steps, albeit at the cost of demanding more qubits with longer coherence times. 

In this work, we focus on time-domain ultrafast optical spectroscopy, where the system is probed by an ultrashort optical pulse.  Using a grid-based discretization approach, we extend the simulation of discrete quantum systems studied in \cite{PhysRevResearch.4.043106, Bruschi2024} to continuous quantum fields, exemplified by the dynamics of electron fields in semiconductors. To reduce the number of qubits required, we adopt a semiclassical approximation for light–matter interactions, in which the matter subsystem is treated fully quantum mechanically, while the electromagnetic field is described classically \cite{haug_koch}. Its influence on the quantum system enters through time-dependent terms in the electronic Hamiltonian that capture induced dipole interactions, thereby neglecting explicit field quantization. We further generalize the framework to the finite-temperature regime. 

Overall, quantum simulations of time-domain spectroscopy are motivated by two key considerations. First, compared to experimental spectroscopy, quantum simulations act as a “time-scale microscope,” mapping ultrafast electronic interactions onto the dynamics of qubit states, which evolve much more slowly in a controllable environment. Second, compared to classical simulations, quantum approaches are expected to exhibit a quantum advantage when many-body effects are taken into account.

The remainder of this paper is organized as follows. Section~II introduces the theoretical framework, including the semiconductor Bloch equations and the Lindblad description of dephasing. Section~III presents the quantum-simulation methodology, including the fermion-to-qubit mapping, Hamiltonian decomposition, implementation of pure dephasing through quantum trajectories, and treatment of finite-temperature effects. Section~IV demonstrates the approach through simulations of linear absorption and gain spectra in GaAs and compares quantum and classical results. Section~V discusses the expected computational advantages, limitations, and relation to other quantum-simulation problems. Finally, Section~VI summarizes the main findings and outlines directions for future research.

\begin{figure}[t]
\centering
\includegraphics[width=0.9\linewidth]{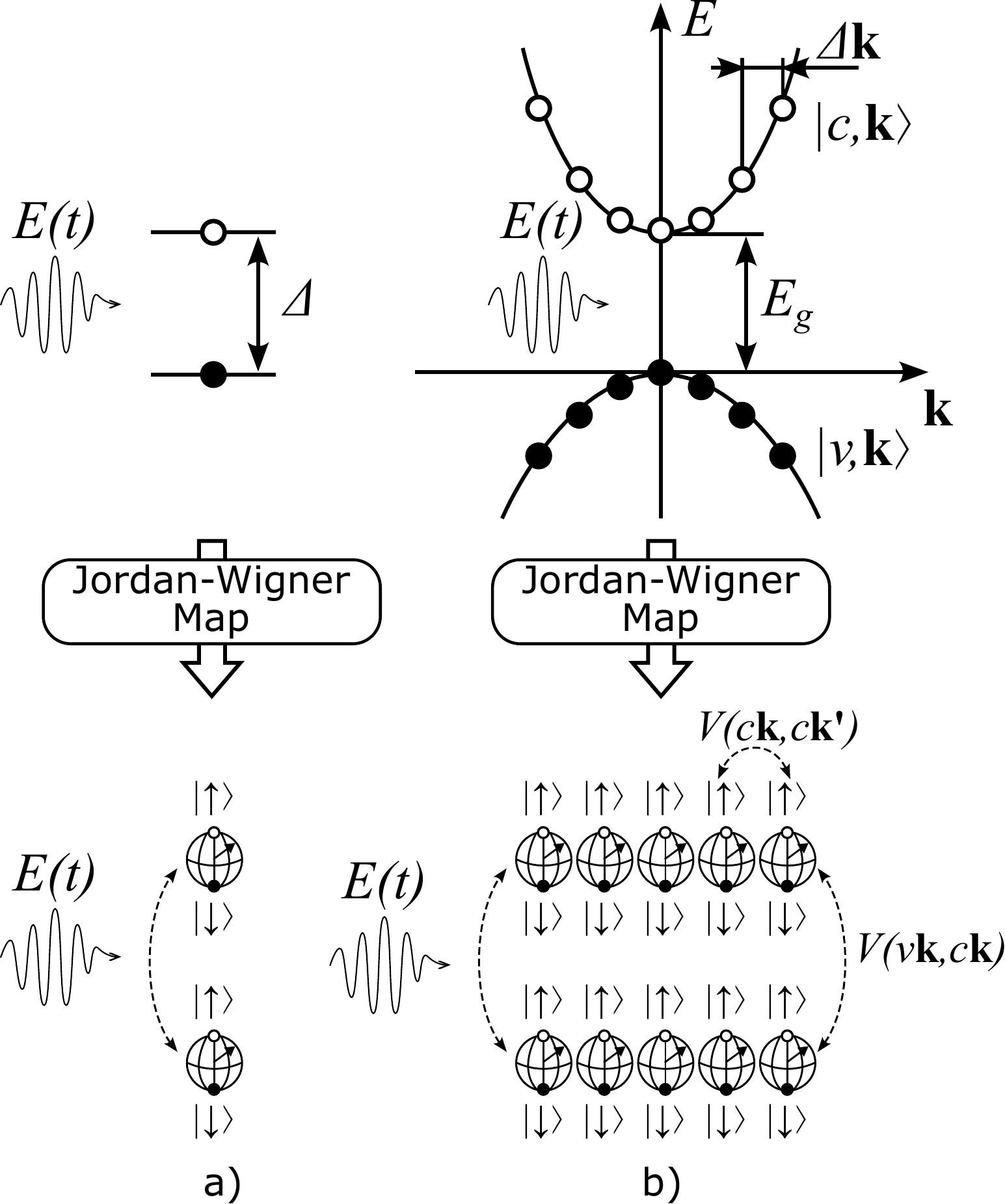}
\caption{\label{fig:states} Mapping spectroscopic experiments onto the problem of Hamiltonian simulation for (a) a two-level system and (b) a direct-band two-band semiconductor. In the semiclassical approximation, light–matter interactions are modeled through time-dependent induced dipole moments and implemented on a quantum computer using parameterized entangling two-qubit gates.
}
\end{figure}

\section{Problem statement}

\subsection{Hamiltonian}

In crystalline semiconductors in the absence of external perturbations, the electron quantum fields are periodic in space, and the quasiparticle energy spectrum, $\varepsilon_{j,\mathbf{k}}$, exhibits a band structure. The corresponding quantum states are labeled by two quantum numbers, $\ket{j, \mathbf{k}}$, where $j$ is the band index and $\mathbf{k}$ is the wave vector confined to the first Brillouin zone (we have omitted the spin quantum number, as magnetic effects are not considered in this work). For optical spectroscopy, where the interaction between a semiconductor and EM fields is the key effect, it is usually sufficient to consider only the two energy bands closest to the Fermi level - one valence band and one conduction band, $j \in \{v, c\}$, separated by an energy gap $E_g$, as is illustrated in Fig.~\ref{fig:states}.

For many-particle systems with interactions, we use the basis set $\ket{j, \mathbf{k}}$ to construct the second-quantization representation of operators and states \cite{Mahan_2015, fetter2012quantum, Coleman_2015}. In this representation, many-particle quantum states are specified by the occupation numbers of single-particle states:

\begin{equation}
    \ket{\Psi} = \ket{ n_{v, \mathbf{k}_1}, n_{c, \mathbf{k}_1},  \dots, n_{v, \mathbf{k}_N}, n_{c, \mathbf{k}_N} },
    \label{wf}
\end{equation}

 For instance, the ground state of a semiconductor is given by the following Slater determinant, describing a fully occupied valence band and an empty conduction band:
\begin{equation}
    \ket{gs} = \ket{101010...}= a_{v, \mathbf{k}_1}^{\dagger} a_{v, \mathbf{k}_2}^{\dagger} \dots a_{v, \mathbf{k}_N}^{\dagger} \ket{0}.
    \label{vac}
\end{equation}

The total Hamiltonian describing the propagation of an electromagnetic pulse in a semiconductor within the second-quantization formalism comprises the kinetic energy of quasiparticles, their interaction with the electromagnetic field, and Coulomb coupling between charged quasiparticles, and is given by \cite{chow1999semiconductor, haug_koch}:
\begin{equation}
    H = H_{s} + H_C + H_{I}(t),
    \label{ham}
\end{equation}
where
\begin{equation}
        H_{s}  = \sum_{j \in \{v,c \}, \mathbf{k}} \varepsilon_{j, \mathbf{k}} a_{j, \mathbf{k}}^{\dagger}a_{j, \mathbf{k}},
        \label{ham1}
\end{equation}
\begin{equation}
        H_C  = \sum_{\substack{i,j \in \{v,c \}, \\ \mathbf{k}, \mathbf{k}', \mathbf{q}\neq 0}} V_{\mathbf{q}}, a_{i, \mathbf{k}-\mathbf{q}}^{\dagger}a_{j, \mathbf{k}+\mathbf{q}}^{\dagger}a_{j, \mathbf{k}}a_{i, \mathbf{k}'}
        \label{ham2}
\end{equation}
\begin{equation}
    H_I(t) = E(t) \sum_{\mathbf{k}} d_{\mathbf{k}}^{cv} a_{c, \mathbf{k}}^{\dagger}a_{v, \mathbf{k}} + d_{\mathbf{k}}^{cv*} a_{c, \mathbf{k}}a_{v, \mathbf{k}}^{\dagger},
    \label{ham3}
\end{equation}
and $\varepsilon_{j,\mathbf{k}}$ denotes the quasiparticle energy of electrons and holes, $j \in \{c, v\}$, $E(t)$ is the applied electromagnetic field, $d_{\mathbf{k}}^{cv}$ is the interband dipole moment, and $V_{\mathbf{q}}$ is the Fourier-transformed Coulomb potential.

\subsection{Observables and spectral characteristics}

The quantum states of open systems at finite temperature are mixed states, which are represented by the density matrix $\rho$. Using the density matrix, we can compute the observable whose time evolution we aim to determine within this framework -- the induced dipole moment or polarization:
\begin{equation}
    \mathbf{P}(t) = \frac{1}{V} \text{Tr} \left(  \widehat{\mathbf{d}} \rho \right),
    \label{pol_trace}
\end{equation}
which is in turn related to the electric susceptibility which reads: 
\begin{equation}
    \mathbf{P}(t) = \int dt' \chi(t-t') \mathbf{E}(t'),
\end{equation}
or in the frequency domain is:
\begin{equation}
    \chi(\omega)  = \frac{\mathbf{P}(\omega)}{\mathbf{E}(\omega)}.
    \label{susceptibility}
\end{equation}
The electric susceptibility enables the calculation of the absorption coefficient, 
$\alpha(\omega)$ \cite{haug_koch}:
\begin{equation}
    \alpha(\omega) = \frac{4 \pi \omega}{n_b c} \text{Im} \left[ \chi(\omega) \right],
    \label{absorption}
\end{equation}
where $n_b$ is background refractive index, and $c$ is the speed of light in vacuum.

In this work, we consider only interband optical processes, while intraband transitions are neglected. In this case, only the single-electron density-matrix elements that are diagonal in the wave vector contribute to the observable in Eq.~\eqref{pol_trace}, which results in the expression:
\begin{equation}
    \mathbf{P}(t) = \frac{1}{V} \sum_{\mathbf{k}} \mathbf{d}_{\mathbf{k}}^{cv} \rho_{\mathbf{k}\mathbf{k}}^{vc}(t) + \mathbf{d}_{\mathbf{k}}^{vc} \rho_{\mathbf{k}\mathbf{k}}^{cv}(t),
    \label{polarization}
\end{equation}
where $\rho_{\mathbf{k}\mathbf{k}'}^{ij}=\langle a_{i, \mathbf{k}}^{\dagger} a_{j, \mathbf{k}'} \rangle$ is an element of the one-particle reduced density matrix and $i,j \in \{v,c\}$. In what follows, we adopt the following notation when discussing systems with decoupled wave vectors:
\begin{equation}
n_{c, \mathbf{k}}=\rho_{\mathbf{k}\mathbf{k}}^{cc}, \qquad n_{v, \mathbf{k}}=\rho_{\mathbf{k}\mathbf{k}}^{vv}, \qquad p_{\mathbf{k}}=\rho_{\mathbf{k}\mathbf{k}}^{vc}.
\end{equation}

\subsection{Time-evolution and dephasing}

The time evolution of the density matrix is given by the Liouville--von Neumann equation: 
\begin{equation}
   \frac{d\rho}{dt} = \mathcal{L} \left( \rho \right), 
   \label{eq:liouville}
\end{equation}
where $\mathcal{L}$ is Liouville superoperator, also known as Liouvillian \cite{Breuer}, given by:
\begin{equation}
       \mathcal{L} \left( \rho \right) = -\frac{i}{\hbar} [H_s, \rho]-\frac{i}{\hbar} [H_C, \rho] -\frac{i}{\hbar} [H_I(t), \rho]+\mathcal{D}\left( \rho \right).
   \label{eq:liouville1}
\end{equation}
The last term in Eq.~\eqref{eq:liouville1}, which accounts for non-Hamiltonian dynamics, describes the interaction between an open quantum system and its environment. This term formally arises from the reduced density matrix formalism, where the environmental degrees of freedom are traced out, leading to integro-differential kinetic equations known as the Nakajima–Zwanzig equation \cite{Nakajima1958, Zwanzig, Breuer}. Explicit evaluation of $\mathcal{D}\left( \rho \right)$ requires approximations in which the system–bath interaction is described either by a microscopic theory or by a phenomenological model. 

The microscopic theory treats the bath as a set of additional fermionic or bosonic degrees of freedom, explicitly coupled to the system, which leads to a hierarchical system of equations of motion \cite{Yoshitaka2020, delgado2024unitary,KIRA2006155}. 

The phenomenological approach is based on the quantum jump method \cite{RevModPhys.70.101}, where the system–environment interaction is modeled as a stochastic, non-unitary and approximate time evolution, containing phenomenological parameters such as relaxation and dephasing times. This approach generally disregards long-time memory effects in the system-bath evolution, an assumption known as the Markov approximation. A common form of such a phenomenological model is given by the Lindblad master equation \cite{Breuer}, in which the interaction with the environment is described by a superoperator known as the dissipator:
\begin{equation}
    \mathcal{D}( \rho) = \sum_{\alpha = (j,  \mathbf{k})}  \left( L_{\alpha} \rho L_{\alpha}^{\dagger} - \frac{1}{2} \left\{ L_{\alpha}^{\dagger} L_{\alpha}, \rho \right\} \right),
    \label{lindblad}
\end{equation}
where $L_{\alpha}$ are jump operators representing interactions with the environment, $\left\{\cdot, \cdot \right\}$ denotes the anticommutator. 

We restrict our consideration to the pure dephasing process, for which the quantum jump operator is
\begin{equation}
    L_{j,  \mathbf{k}} = \sqrt{\gamma} a_{j, \mathbf{k}}^{\dagger} a_{j, \mathbf{k}},
    \label{eq:jump}
\end{equation}
where $\gamma$ is the inverse dephasing rate, leading to (in the element-wise form):
\begin{equation}
        \left[\mathcal{D}(\rho) \right]_{\mathbf{k}', \mathbf{k}''}^{n, m} = \gamma \rho_{\mathbf{k}', \mathbf{k}''}^{n, m} \left(\delta_{n,m}\delta_{\mathbf{k}', \mathbf{k}''} - 1 \right)
        \label{dissipator}
\end{equation}

Under this approximation and when $H_C=0$, Eq.~\eqref{eq:liouville} in the element-wise form yields a closed system of independent ordinary differential equations commonly referred to as the semiconductor Bloch equations \cite{haug_koch, 10.1063/1.121140, PhysRevB.103.115203}:
\begin{equation}
          \frac{d p_{\mathbf{k}}}{dt} = i \omega_{\mathbf{k}} p_{\mathbf{k}} -i \omega_{\mathbf{k}}^R \left(n_{c,\mathbf{k}} - n_{v,\mathbf{k}} \right)  -
          \gamma p_{\mathbf{k}},   
    \label{sbe1}
\end{equation}
\begin{equation}
   \frac{d n_{c,\mathbf{k}}}{dt} =- 2\text{Im} \left( \omega_{\mathbf{k}}^R p_{\mathbf{k}}^* \right),  
   \label{sbe2}
\end{equation}
\begin{equation}
   \frac{d n_{v,\mathbf{k}}}{dt} = 2\text{Im} \left( \omega_{\mathbf{k}}^R p_{\mathbf{k}}^* \right).
   \label{sbe3}
\end{equation}
where $\omega_{\mathbf{k}} = \left( \varepsilon_{c,\mathbf{k}} - \varepsilon_{v,\mathbf{k}} \right) / \hbar$ is the transition frequency, and $\omega_{\mathbf{k}}^R = d_{\mathbf{k}}^{cv} E(t) / \hbar$ is the Rabi frequency.

Incorporating the Coulomb coupling, $H_C$, gives rise to four-operator products of creation and annihilation operators on the right-hand side of the semiconductor Bloch equations. The exact closed-form representation of these higher-order operators is, in general, not known \cite{KIRA2006155}. To close the system, several approaches can be taken. The simplest consists of incorporating $H_C$ into $\mathcal{D}( \rho)$, assuming that the effect of scattering due to many-body interactions can be reproduced by introducing a bath of bosonic or fermionic degrees of freedom and their corresponding couplings. For instance, dephasing, as described above, can result from electron-electron interactions. This approach provides an approximate phenomenological model of scattering-assisted broadening.

A more rigorous treatment of $H_C$ requires deriving additional kinetic equations for higher-order correlations. In this case, inclusion of progressively higher-order correlations leads to a large, or even infinite, hierarchy of coupled kinetic equations, known as the hierarchy problem \cite{KIRA2006155}. This hierarchy can be truncated using approximations such as the Hartree-Fock approximation \cite{chow1999semiconductor, KIRA2006155}, which factorizes four-operator products into products of two operators, thereby neglecting higher-order correlations \cite{haug_koch, KIRA2006155}. The Coulomb interaction within the Hartree-Fock approximation leads to a renormalization of the electron and hole energies and a redefinition of the Rabi frequency in Eqs.~\eqref{sbe1}-\eqref{sbe3}:
\begin{equation}
    \varepsilon_{j,\mathbf{k}}' = \varepsilon_{j,\mathbf{k}} + \sum_{\mathbf{q}} V_{|\mathbf{k} - \mathbf{q}|} n_{j, \mathbf{q}}, \qquad j \in \{v, c\}
    \label{HF_renorm}
\end{equation}
\begin{equation}
    \omega_{\mathbf{k}}^R = \frac{d_{\mathbf{k}}^{cv} E(t)}{\hbar} + \frac{1}{\hbar} \sum_{\mathbf{q} \neq \mathbf{k}} V_{|\mathbf{k} - \mathbf{q}|} p_{\mathbf{q}} .
    \label{HF_Rabi}
\end{equation}

Analyzing Eqs.~\eqref{sbe1}--\eqref{sbe2}, we observe that, in the two-band system, for each wave vector in the discretized Brillouin zone the equations are formally identical to the two-level optical Bloch equations, as illustrated in Fig.~\ref{fig:states}. Without Coulomb interactions, each two-level system is independent, whereas inclusion of Coulomb interactions couples them. Thus, methods developed for quantum simulation of two-level systems can be extended to the optical response of direct-bandgap semiconductors.

The goal of this work is to solve the system of Eqs.~\eqref{sbe1}–\eqref{sbe3} on a quantum computer under various approximations.

\subsection{Dimensionality and axial approximation}

The computation of optical spectral characteristics, whether on classical or quantum computers, can be simplified by reducing the problem to a one-dimensional case under the axial approximation. This approximation neglects the angular dependence of electron and hole transport by assuming that the medium and, consequently, the band structure are isotropic \cite{1518141}. In this case, results obtained for a one-dimensional semiconductor can be used to evaluate the optical properties of semiconductors with arbitrary dimensionality.

This can be achieved by rewriting the sum in Eq.~\eqref{polarization} in the continuous limit using polar coordinates \cite{haug_koch}:
\begin{equation}
    \sum_{\mathbf{k}} = \sum_{\mathbf{k}} \frac{\Delta k}{\Delta k} \rightarrow \left( \frac{L}{2\pi} \right)^D \int d \Omega_D \int\limits_0^{k_{max}} dk \; k^{D-1},
    \label{continous_limit}
\end{equation}
where $D=1,2,3$ is dimensionality, $L$ is the generalized volume of the system, $\Omega_D$ is is the space angle in $D$ dimensional space, and $k$ is the length of the wave vector.

Within the axial approximation, the solid-angle integration in Eq.~\eqref{continous_limit} can be performed analytically:
\begin{equation}
    \sum_{\mathbf{k}} \rightarrow \left( \frac{L}{2\pi} \right)^D \frac{2 \pi^{D/2}}{\Gamma(D/2)} \int\limits_0^{k_{max}} dk \; k^{D-1},
    \label{axial_approximation}
\end{equation}
where $\Gamma(\cdot)$ is the Gamma function. 

With the axial approximation, the problem reduces to a one-dimensional problem. In this approach, it is sufficient to compute the polarization on a one-dimensional grid, as illustrated in Fig.~\ref{fig:states}, and subsequently apply Eq.~\eqref{axial_approximation} to describe systems of arbitrary dimensionality.

The axial approximation breaks down for band structures with strong anisotropy. In this case, the entire 2- or 3-D k-space must be discretized on a grid, which increases the number of sampling points and, consequently, the number of qubits required for quantum simulations.

\section{Simulation approach}

\subsection{Hamiltonian simulation}

The problem stated in the previous section shows that accurate simulation of ultrafast spectroscopy involves several physical effects that lie beyond the unitary dynamics native to qubit-based quantum computers, posing the challenge of efficient implementation within quantum computing constraints. Another challenge is that spectroscopic analysis requires access to dynamical information at a large number of time steps and over sufficiently long time intervals, as these determine the temporal resolution of the spectroscopic technique. Lastly, the fact that electronic fields in crystalline semiconductors are continuous - although periodic - poses a challenge for their simulation on digital quantum computers. However, discretization of these fields is also required for simulations on classical digital computers, where the corresponding methodologies are well established. 

The standard discretization approach relies on the \textit{Born–von Karman boundary conditions}. These conditions imply that, instead of treating the semiconductor as an infinite system, one considers a crystal of finite size (defined by $n$ translations of primitive cell defined by the primitive translation vectors $\mathbf{a}_j$) and applies periodic boundary conditions at its borders: $\Psi(\mathbf{r} + n \mathbf{a}_j, t) = \Psi(\mathbf{r}, t)$. This converts the continuous Brillouin zone into a grid of wave vectors, with the spacing $\Delta \mathbf{k} = 2 \pi / (n \mathbf{a}_j) $ as is illustrated in Fig.~\ref{fig:states}. Thus, the combined effect of the crystallographic periodicity and the Born–von Karman boundary conditions results in a finite set of allowed wave vectors. 

We can further reduce this set by considering only wave vectors in the vicinity of the high-symmetry points of the Brillouin zone and by including only a limited number of energy bands. This region of wave vectors and reduced energy window is where the most light-matter interactions of interest occur, analogous to the concept of an active space in quantum chemistry \cite{helgaker2013molecular}. Electrons occupying states outside this window can be considered "frozen", as they do not change their states under irradiation by the EM field with given parameters. The precise wavevector and energy domain boundaries are determined by the parameters of EM fields - frequency and amplitude (linear or nonlinear regime).

The fermionic creation and annihilation operators introduced above (see Eqs.~\eqref{ham1}–\eqref{ham3}) can be simulated on a quantum computer by imposing the fermionic anti-commutation relations on qubits that obey a tensor-product spin algebra.
To simulate fermionic systems on qubits, a fermion-to-qubit mapping, as is illustrated in Fig. \ref{fig:states}, is required \cite{Chien2026}.
A wide range of fermion-to-qubit mappings has been developed \cite{chen2024error, miller2023bonsai, li2022unified, whitfield2011simulation, nielsen2005fermionic, somma2002simulating, seeley2012bravyi, bravyi2002fermionic} and in this work, we use the Jordan--Wigner transformation (JWT) \cite{jordan1928pauli}, the earliest and most widely used mapping scheme. 

In the JWT, each fermionic mode is mapped to a single qubit, with occupations encoded as qubit states, so that $|0\rangle$ represents an empty orbital and $|1\rangle$ a filled one. The key challenge in the mapping is preserving the fermionic sign factor 
$\eta_j(n) = (-1)^{\sum_{k<j} n_k}$, which encodes the anti-commutation relations.
In the JWT, this is accomplished by introducing a string of Pauli-$Z$ operators acting on all qubits with index $k<j$: $Z_1 Z_2 \cdots Z_{j-1}$. This operator has eigenvalues $\pm 1$ depending on the parity of occupied modes below $j$, 
exactly reproducing the phase of a fermionic operator. 

With this parity structure in place, the fermionic creation and annihilation operators are mapped to Pauli operators as:
\begin{align}
    a_j 
        &\;\longrightarrow\; 
        \left( Z_1 Z_2 \cdots Z_{j-1} \right)
        \frac{X_j + i Y_j}{2}, \label{jwt1} \\
    a_j^\dagger 
        &\;\longrightarrow\; 
        \left( Z_1 Z_2 \cdots Z_{j-1} \right)
        \frac{X_j - i Y_j}{2}. \label{jwt2}
\end{align}
With these transformations, the fermionic number operator is mapped to the Pauli operator:
\begin{equation}
    a_j^\dagger a_j 
    \quad\longrightarrow\quad 
    \frac{1 + Z_j}{2},
    \label{jwt1_1}
\end{equation}
\begin{equation}
    a_j a_j^\dagger 
    \quad\longrightarrow\quad 
    \frac{1 - Z_j}{2}.
    \label{jwt2_1}
\end{equation}

The Pauli operators $X_j$ and $Y_j$ flip the $j$th qubit between its occupied and unoccupied states, 
while the $Z$-string enforces the correct fermionic phase.
Using this mapping, any fermionic Hamiltonian $H$ can be expressed as a linear combination of multi-qubit Pauli strings,
\begin{equation}
    H = \sum_\ell h_\ell\, P_\ell, 
    \qquad P_\ell = \sigma_{i_1}^{(1)} \otimes \sigma_{i_2}^{(2)} 
        \otimes \cdots \otimes \sigma_{i_N}^{(N)},
        \label{matrix_basis}
\end{equation}
where $\sigma_{i_j}^{(j)} \in \{ I, X, Y, Z \}$. This qubit Hamiltonian forms the basis for implementing Hamiltonian simulation methods. The representation in Eq.~\eqref{matrix_basis} holds for any quantum-mechanical operator, since Pauli matrices (and their tensor products or so-called strings) form a complete basis for the Hilbert–Schmidt space.

Note that JWT is not the most efficient method for mapping fermionic operators onto qubits. The implementation of long Pauli strings requires deep quantum circuits, with the number of gates scaling as $\mathcal{O}(N)$, where $N$ is the number of sites. More efficient alternatives include the Bravyi-Kitaev encoding, the Bravyi--Kitaev superfast encoding, and others~\cite{BRAVYI2002210, PhysRevA.95.032332, PhysRevB.104.035118, Clinton2021}. However, since the present work is intended primarily as a proof of concept, we employ JWT since it is conceptually the simplest method.

\begin{figure}[!t]
    \centering
\includegraphics[width=0.7\linewidth]{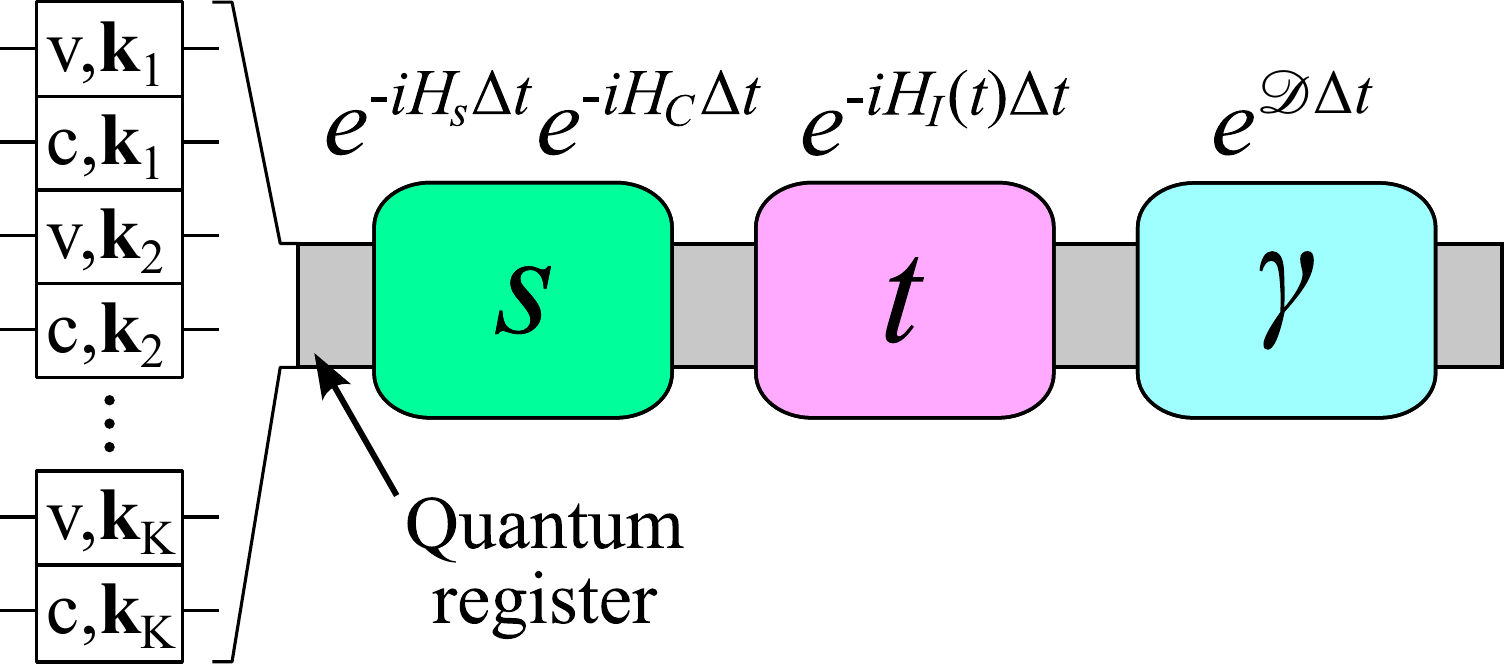}
    \caption{Quantum circuit diagram for one Trotter step of the Hamiltonian simulation}
    \label{fig:trotter_spep}
\end{figure}

After discretizing the Brillouin zone and mapping the Hamiltonian to a qubit representation, the problem reduces to computing time-dependent expectation values - such as the polarization or induced dipole moment in Eq.~\eqref{polarization} - on a quantum computer. The expectation values depend on the time evolution of the density matrix elements. To obtain the time evolution of the density matrix elements on a quantum computer, it is convenient to employ the Schr\"odinger picture together with standard \textit{Hamiltonian simulation} techniques. This approach involves preparing the state $\ket{\psi(t)} = U \left(t, t_0 \right) \ket{\psi(0)},$ where
\begin{equation}
    U \left(t, t_0 \right) = \mathcal{T} \left[ e^{-i\int_{t_0}^t dt' H(t')} \right].
    \label{eq:u}
\end{equation}
This state can be use to evaluate the expectation value of a Hermitian operator, $\langle \hat{O}  \rangle = \bra{\psi(t)} \hat{O} \ket{\psi(t)}$. Both unitary time-evolution simulation and expectation-value evaluation are considered standard tasks for quantum computers, and we discuss their implementation details further below. First, however, we establish how the information obtained from these tasks can be related to the quantity of interest: the time evolution of the density matrix. An element of the density matrix in the computational basis at time $t$ is given by:
\begin{equation}
    \begin{split}
    \left[ \rho(t) \right]_{\mathbf{k},\mathbf{k}'}^{nm} &= \text{Tr} \left( \rho(t) a_{n,\mathbf{k}}^{\dagger} a_{m,\mathbf{k}'} \right) \\
    &=\text{Tr} \left( U\left(t,t_0 \right) \rho(t_0) U^{\dagger}\left(t_0,t \right) a_{n,\mathbf{k}}^{\dagger} a_{m,\mathbf{k}'} \right),
    \end{split}
    \label{eq:dyn}
\end{equation}
where $\rho(t_0)$ is the density matrix of the initial state, which, in the general case of mixed states, can be represented in the form
\begin{equation}
    \rho(t_0) = \sum_j w_j \ket{\psi_j(t_0)} \bra{\psi_j(t_0)}.
\end{equation}
Substituting this expression into Eq.~\eqref{eq:dyn}, and using the cyclic property of the trace together with 
$\ket{\psi_j(t)} = U(t,t_0)\ket{\psi_j(t_0)}$, we obtain:

\begin{equation}
    \left[ \rho(t) \right]_{\mathbf{k},\mathbf{k}'}^{nm} =   \sum_j w_j \bra{\psi_j(t)} a_{n,\mathbf{k}}^{\dagger} a_{m,\mathbf{k}'} \ket{\psi_j(t)} ,
    \label{eq:dyn1}
\end{equation}
The statistical weights $w_j$ encode the initial conditions of $\rho(t_0)$ as well as the ensemble average over randomly generated circuits in the quantum trajectory method. The obtained Eq.~\eqref{eq:dyn1} provides a framework for conceptualizing ultrafast spectroscopy simulations on quantum computers.

The operator $a_{n,\mathbf{k}}^{\dagger} a_{m,\mathbf{k}'}$ in Eq.~\eqref{eq:dyn1} is non-Hermitian, its expectation value on a quantum computer can be evaluated in two separate runs, corresponding to its real and imaginary parts. Specifically, since we consider only vertical optical transitions, we evaluate the expectation values of the following Hermitian operators with equal wave vector indices - only these contribute to the macroscopic polarization in Eq.~\eqref{polarization}:
\begin{equation}
\hat{O}_{\mathbf{k}}^{\text{Re}} = a_{n,\mathbf{k}}^{\dagger}a_{m,\mathbf{k}} + a_{m,\mathbf{k}'}^{\dagger}a_{n,\mathbf{k}},
\label{re}
\end{equation}
\begin{equation}
\hat{O}_{\mathbf{k}}^{\text{Im}}= i a_{n,\mathbf{k}}^{\dagger}a_{m,\mathbf{k}'} - i a_{m,\mathbf{k}}^{\dagger}a_{n,\mathbf{k}},
\label{im}
\end{equation}
from which the desired complex expectation value can be reconstructed. As a result of JWT, each of these Hermitian operators can then be expressed as a linear combination of two unitary operators represented by Pauli strings (see  Eq.~\eqref{matrix_basis}):
$$\hat{O}_{\mathbf{k}}^{\text{Re}} = \frac{1}{2} \left(YY + XX \right), \quad \hat{O}_{\mathbf{k}}^{\text{Im}}= \frac{1}{2} \left(XY - YX \right).$$
 Consequently, the evaluation of Eq.~\ref{eq:dyn1} requires four additional measurement circuits at each time point. For $N_{\mathrm{shots}}$ shots per circuit, the number of measurements is  $4N_{\mathrm{shots}}$ shots per time point. In this setup, the measurements corresponding to different wave vectors are performed in parallel.

The time-dependent Hamiltonian simulations on quantum computers, aimed at representing and evaluating of the operator $U(t,t_0)$ in Eq.~\eqref{eq:u} by means of quantum circuits, has attracted considerable attention in recent years and has led to substantial progress in the development of quantum algorithms. Representing the Hamiltonian exponentiation in \(U(t,t_0)\) is a trivial task if the Hamiltonian is expressed as a single Pauli string in Eq.~\eqref{matrix_basis}. The problem arises when the Hamiltonian is a sum of non-commuting terms, which is most often the case. This is where approximate methods, such as the Trotter product formula, linear combination of unitaries \cite{PRXQuantum.6.010359}, quantum signal processing \cite{PhysRevLett.118.010501, PRXQuantum.5.020368, PhysRevA.110.012612} or qubitization \cite{low2019hamiltonian}, need to be applied. This variety of algorithms has been developed as a result of efforts to improve efficiency, which is largely determined by the accumulation of errors during time propagation and by the number of quantum gates required to implement the evolution operator. Each algorithm offers different trade-offs between circuit depth, accuracy, and the number of ancillary qubits required. Algorithms such as qubitization \cite{low2019hamiltonian} and quantum signal processing \cite{PhysRevLett.118.010501, PRXQuantum.5.020368, PhysRevA.110.012612} are optimal in terms of gate count; however, they require additional ancillary qubits and a significant number of two-qubit entangling gates. 

In this work, we employ the simpler methods based on Suzuki–Trotter product-formula decompositions, as achieving optimal algorithmic performance is beyond the scope of this study. In addition to its conceptual simplicity, this method offers several practical advantages: it does not require ancillary qubits, naturally discretizes the time interval, and can be straightforwardly applied with an arbitrary number of time steps. The latter feature is particularly important for our purposes, as it enables the estimation of spectroscopic characteristics at intermediate times during the evolution.

 To approximate the time propagation $\ket{\psi_j(t)} = U(t,t_0)\ket{\psi_j(t_0)}$, we discretize the simulation time into
\(m = t/\Delta t\) steps and apply the first-order Suzuki--Trotter formula, \cite{PhysRevLett.106.170501, suzuki1985decomposition}
\begin{widetext}
\begin{equation}
    U \left(t, 0 \right) = \mathcal{T} \left[ e^{-i\int_{0}^t dt' H(t')} \right] 
    \;\approx\;
    \prod_{j=1}^{m}
        e^{-i H_{\mathrm{s}}\, \Delta t}
        e^{-i H_{\mathrm{C}}\, \Delta t}
        e^{-i H_{I}(j\Delta t)\, \Delta t}
        e^{-i H_{\mathcal{D}}\, \Delta t},
    \label{eq:trotter}
\end{equation}
\end{widetext}
where \(t_j = j\Delta t\) denotes the time at the $j$th step, $H_{\mathcal{D}}$ is an auxiliary Hemritian operator, resulting from the quantum trajectory method and a stochastic Hamiltonian, that serves as the generator of a unitary evolution constituting one of the trajectories, each sampled with a certain probability, such that averaging over the trajectories recovers the dissipative, non-Hamiltonian dynamics.

Each exponential in Eq.~\eqref{eq:trotter} corresponds to a unitary rotation generated by a Pauli string, which can be straightforwardly implemented on most of the existing quantum computer instruction set architectures. At the level of quantum circuit diagram, the product of these unitary transformations implies that the corresponding gates are applied sequentially, as is illustrated in Fig.~\ref{fig:trotter_spep}. In this diagram, the operations corresponding to the terms $H_{\mathrm{s}}$ and $H_{\mathrm{C}}$ are grouped because both are time-independent. The former contains only single-qubit gates, whereas the latter contains two-qubit gates. 

The unitary operations associated with the external-field interaction term $H_{I}(t_j)$ represent parameterized quantum gates whose rotation angles vary at each Trotter step. 

The operator $H_{\mathcal{D}}$  contributes to the dissipator $e^{\mathcal{D}\, \Delta t}$, which describes non-unitary time evolution. Its implementation on a quantum computer relies on randomized quantum gates that vary from step to step, thereby introducing stochasticity, as discussed in the next section.

Time-domain spectroscopy and its simulations yield values of spectral characteristics for many time steps, determined by the resolution of the experimental setup. For quantum simulations, it is essential to access information at every time step to obtain full information about the system's dynamics: smaller intervals between steps enable coverage of a wider spectral domain, while increasing the number of time steps enhances spectral resolution. This can be achieved by preparing and executing $N$ quantum circuits corresponding to $N$ time intervals: 
\begin{equation}
\Delta T_{j} \in \left\{ \left[ t_0,j\Delta t \right]  \; \vert \; j = 1 \dots N \right\}.
\end{equation}

\begin{figure}[t!]
    \includegraphics[width=0.99\linewidth]{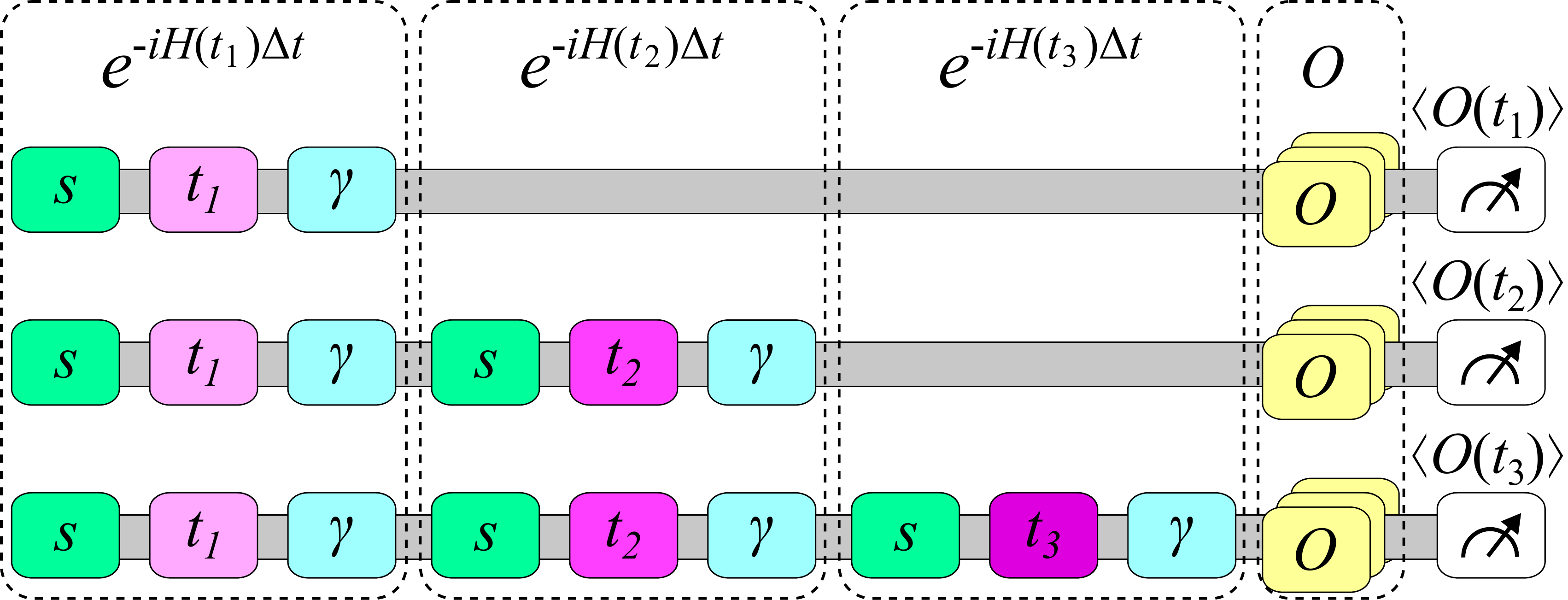}
    \caption{Example quantum circuit for semiconductor time-domain spectroscopy simulations for $N=3$ time steps. The gray ribbon denotes the quantum register of size $2K$, where $K$ is the number of wave vectors in the discretized Brillouin zone, and the factor of two corresponds to the number of energy bands.}
    \label{fig:steps}
\end{figure}

The combined circuits are shown in Fig.~\ref{fig:steps}. Each elementary step has the structure illustrated in Fig.~\ref{fig:trotter_spep} and represents a single Trotter step for the qubit Hamiltonian given by Eq.~\eqref{eq:trotter}. For each time step, the color code denotes the following: 1) the unitary operations in the green block remain unchanged; 2) the parameters of the unitary operations in the pink block change deterministically; 3) operations denoted by the cyan block change stochastically according to a specified distribution function; 4) the yellow blocks in Fig.~\ref{fig:steps} represent a linear combination of unitaries used to represent a non-unitary observable. In this work, the observables are defined by Eqs.~\eqref{re}-\eqref{im}.

As is illustrated in Fig.~\ref{fig:steps}, expectation values at different time points are independent, and quantum information is not reused across steps; to extract information for each time step, the algorithm is restarted from the beginning. While not optimal, this approach avoids ancillary qubits and entangling gates. With sufficient qubits and coherence time, intermediate results can instead be extracted via the Hadamard test~\cite{jens_hadamard}, enabling reuse of quantum information across time steps. This alternative may be more attractive for future quantum computing platforms.

\subsection{Pure phenomenological dephasing and quantum trajectories}

In Section II, we briefly discussed that classical simulations of open systems and many-body interactions may be carried out using either a rigorous microscopic theory or phenomenological stochastic models. Both approaches can be transferred to the realm of quantum simulations.

The microscopic approach to the quantum simulation of open systems is grounded on the fact that any mixed state admits a purification in an enlarged Hilbert space obtained by introducing ancillary degrees of freedom \cite{schlimgen2022quantum, delgado2024unitary}. Within this framework, dissipative dynamics are modeled by explicitly incorporating ancilla qubits that represent the environment, while system–environment interactions are implemented through entangling two-qubit operations between system and ancillary registers. The resulting unitary evolution in the extended Hilbert space reproduces the effective nonunitary dynamics of the reduced system after tracing out the ancilla degrees of freedom. The microscopic approach enables the treatment of non-Markovian dynamics and non-classical correlations \cite{Chen2018, Chen2019, Discord} in system-environment interactions. 

In contrast, the phenomenological approach, which we employ in this work, models system-environment interactions as stochastic processes \cite{Fresch2026} and does not require extra ancillary qubits. Its implementation on quantum computers is based on quantum trajectory theory \cite{10.1063/5.0323944}, which enables the solution of stochastic Schrödinger equations and the simulation of open quantum systems \cite{Fresch2026} by sampling an ensemble of systems undergoing unitary dynamics and averaging over their trajectories. At the circuit level, the trajectories are obtained from repeated circuit executions with variable quantum gates sampled according to a prescribed probability distribution. In the limit of a sufficiently large number of shots, the ensemble-averaged measurement statistics converge to the desired nonunitary dynamics of the open quantum system. System–environment interactions that can be simulated in this manner are referred to as classical. In contrast, correlations that cannot, in principle, be accounted for within this framework are classified as quantum correlations \cite{Chen2018}, with quantum discord serving as their quantitative measure \cite{Discord}.

In this work, we assume that the system and the environment are classically correlated and focus exclusively on phenomenological pure dephasing in Markov limit, as described by Eq.~\eqref{sbe1}. This type of dephasing arises from the Lindbladian superoperator, Eq.~\eqref{lindblad}, with the jump operator given by Eq.~\eqref{eq:jump}. Using JWT given by Eqs.~\eqref{jwt1}--\eqref{jwt2} and neglecting intraband processes, the Lindbladian expressed in terms of Pauli strings reads:

\begin{equation}
    \mathcal{D}(\rho) =  \frac{\gamma}{4} \sum_{j,\mathbf{k}} \left(Z_{j,\mathbf{k}} \rho Z_{j,\mathbf{k}} - \rho \right).
    \label{superoperator}
\end{equation}
where $$Z_{j,\mathbf{k}}=I^{\otimes (\alpha-1)}  \otimes Z \otimes I^{\otimes (2K - \alpha)}$$ with $Z$ denoting the single-qubit Pauli-$Z$ gate. Here, $\alpha$ is the index of the qubit corresponding to the fermionic mode labeled by the quantum numbers $(j,\mathbf{k})$, and $2K$ is the total number of qubits.

The superoperator~\eqref{superoperator} generates a one-parameter semigroup of completely positive and trace-preserving maps. Its finite-time evolution over a step \(\Delta t\) is given by
\begin{equation}
    e^{\mathcal{D}\,\Delta t}(\rho) = \left[ \prod_{j, \mathbf{k}} \left[ (1-p)\, \mathcal{I} + p\, \mathcal{Z}_{j,\mathbf{k}} \right] \right] (\rho) ,
\end{equation}
where $\mathcal{I}$ is the identity channel, $\mathcal{Z}_{j,\mathbf{k}} (\rho ) = Z_{j,\mathbf{k}} \rho Z_{j,\mathbf{k}}$ is a phase-flip channel, $\Pi$ denotes a product of commuting quantum channels, and $p$  is the probability given by
\begin{equation}
    p = \frac{1}{2}\left(1 - e^{-\gamma \Delta t /2} \right).
\end{equation}

Using the quantum trajectory method \cite{Fresch2026,10.1063/5.0323944}, the non-unitary propagation $e^{\mathcal{D}(\rho)\Delta t}$, which describes pure dephasing, can be realized as an ensemble-averaged dynamics corresponding to a Z-gate sampling procedure. At each time interval $\Delta t$, a Z gate is applied to each qubit with probability $p$. The ensemble average over the resulting quantum circuits reproduces the effective time evolution governed by the semiconductor Bloch equations with a phenomenological dephasing rate $\gamma$, as given in Eqs.~\eqref{dissipator}-\eqref{sbe1}.

\subsection{Simulations for finite temperature in thermodynamic quasi-equilibrium}

Spectroscopic experiments are generally sensitive to temperature. Quantum simulations of pure dephasing, as described in the previous section, capture the Lorentzian broadening of spectral features, but they do not adequately describe thermalization processes and, consequently, do not lead to equilibration with the thermal bath. To properly capture thermal effects, the simulations must include a thermal bath model with an appropriate system–bath coupling that ensures thermalization. The initial state should be set as the Gibbs state corresponding to the grand-canonical ensemble. In the general case, preparation of the Gibbs state requires computationally intensive methods, as discussed in \cite{chowdhury2020variational, rajakumar2026gibbs, Chen2025}. These complications can be bypassed and Gibbs state sampling can be avoided when the system is known to remain in equilibrium or near equilibrium, i.e., within the linear optical regime, where the populations remain constant in time. In this case, Eqs.~\eqref{sbe2} and~\eqref{sbe3} do not contribute to the final result, since $n_{j,\mathbf{k}}=const$, and  Eq.~\eqref{sbe1} can be formally integrated as:

\begin{equation}
    \begin{split}
          p_{\mathbf{k}}(t) &= \left[n_{c,\mathbf{k}}(t_0) - n_{v,\mathbf{k}}(t_0) \right] \int\limits_0^t \omega_{\mathbf{k}}^R(t') e^{(i\omega - \gamma)(t-t')} \\
          &= \left[n_{c,\mathbf{k}}(t_0) - n_{v,\mathbf{k}}(t_0) \right]p_{\mathbf{k}}^G(t)
    \end{split}
    \label{sbe_analytical}
\end{equation}
where
\begin{equation}
n_{j,\mathbf{k}}(t_0) = \frac{1}{e^{\beta\left( \varepsilon_{j,\mathbf{k}} - \mu\right)}+1}, \qquad i, j \in \{v, c \}.
\end{equation}

The factor in square brackets is time independent and contains the entire temperature dependence of the solution. At $T=0$ K, it is equal to unity, and we denote the corresponding microscopic polarization by $p_{\mathbf{k}}^{G}(t)$. Consequently, Eq.~\eqref{sbe_analytical} implies that it is sufficient to solve the problem for the ground state. The finite-temperature solution can then be obtained by multiplying the ground-state result by $\left[n_{c,\mathbf{k}}(t_0)-n_{v,\mathbf{k}}(t_0)\right]$, evaluated at the desired temperature and Fermi level. Therefore, instead of explicitly sampling the Gibbs state on a quantum computer, in the linear case we can initialize all qubits such that the system is in the ground state, given by Eq.~\eqref{vac}, and obtain results for any desired temperature by post-processing the quantum-computer outputs. Specifically, the output polarization for each wave vector is weighted by the appropriate factor $\left[n_{c,\mathbf{k}}(t_0) - n_{v,\mathbf{k}}(t_0) \right]$.

Note that the distribution functions $n_{c,\mathbf{k}}(t_0)$ and $n_{v,\mathbf{k}}(t_0)$ may be characterized by different temperatures and Fermi levels; this corresponds to a quasi-equilibrium state. In general, a rigorous treatment of non-equilibrium dynamics requires solving the full quantum kinetic equations for the many-body density matrix, a task that is generally intractable. The steady-state quasi-equilibrium approximation avoids this difficulty by assuming that carrier--carrier scattering is much faster than carrier generation, recombination, and external driving. Consequently, electrons and holes rapidly thermalize within their respective bands and may be described by separate Fermi--Dirac distributions characterized by their own quasi-Fermi levels.

\subsection{Resource requirements and implementation details}

To demonstrate the proposed approach, we perform simulations using the IBM Qiskit quantum simulation environment and compare the results with classical simulations. Specifically, the simulations employed the Qiskit-Aer simulator with high-performance computing (HPC) support on x86-64 CPU clusters. The computations were parallelized using the Message Passing Interface (MPI) functions.

The number of qubits required for the two-band spectroscopy simulations is $2K$, where $K$ is the number of simulated wave vectors. The total number of quantum gates scales linearly with the number of time steps (Trotter steps), $N$, and can be expressed as:

\begin{equation}
N_{\mathrm{gates}}
=
K\left[
\left(\alpha+\beta+\nu \right)N+n_0
\right],
\label{num_gates}
\end{equation}
where  $\alpha$ and $\beta$ denote the numbers of two-qubit and single-qubit gates, respectively, applied per time step and per wave vector. The parameter $\nu$ denotes the number of placeholders for randomly sampled single-qubit gates per time step and per wave vector. Finally, $n_0$ is the number of additional single-qubit gates required to prepare the initial population of the system.

\begin{figure}
    \centering
    \includegraphics[width=0.9\columnwidth]{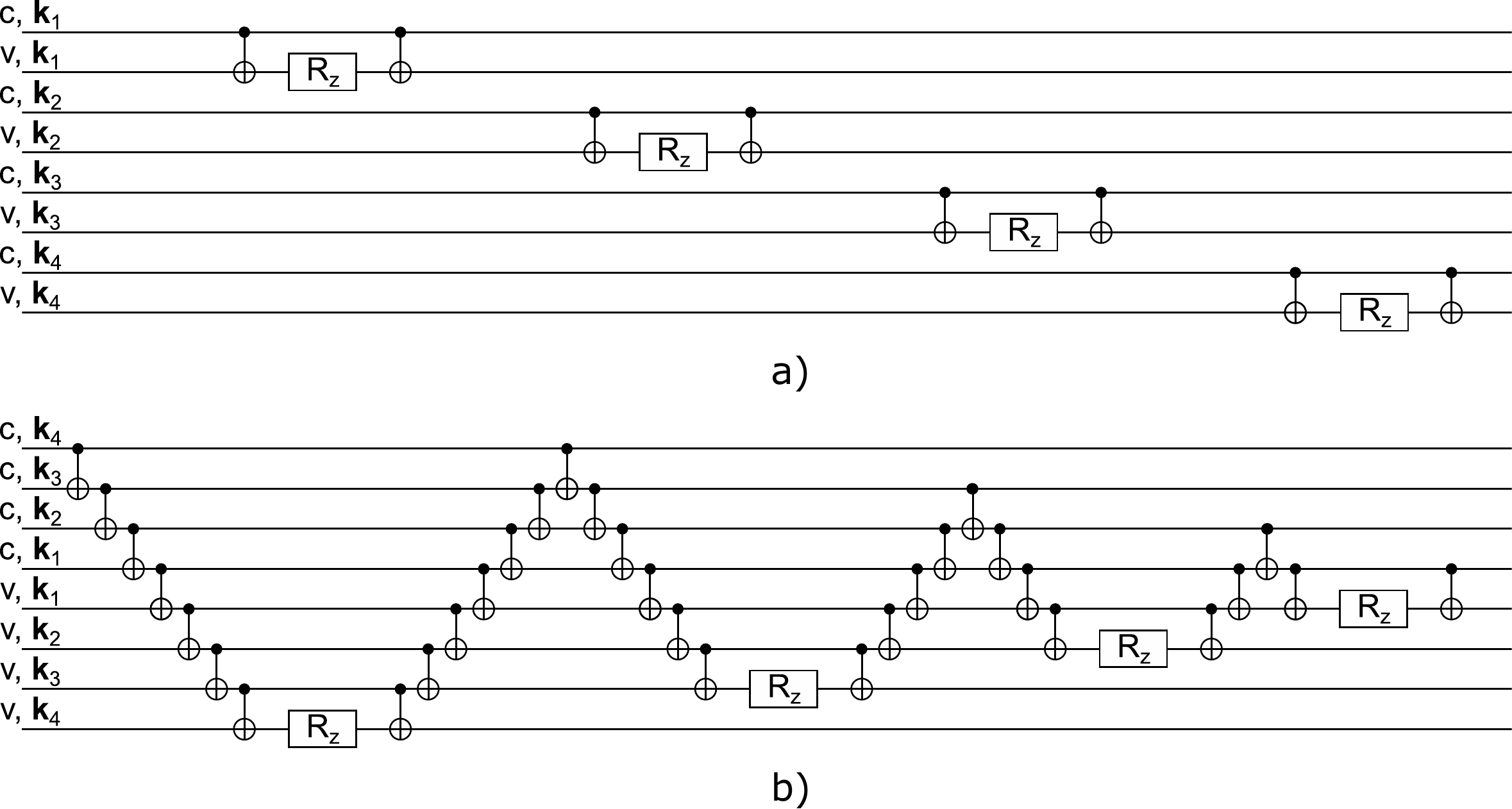}

\caption{Result of the Jordan--Wigner transformation, Eqs.~\eqref{jwt1}--\eqref{jwt2}, applied to the operator $E(t)\sum_{\mathbf{k}} d_{\mathbf{k}}^{cv}a_{c,\mathbf{k}}^{\dagger}a_{v,\mathbf{k}}$ in Eq.~\eqref{ham3}, for two different orderings of the quantum numbers. The rotation $R_z$ is determined by the phase of the signal $E(t)$.}
    \label{fig:orderings}
\end{figure}

The number of two-qubit gates, $\alpha$, is determined by the type of interactions in the system and by the chosen ordering of the quantum numbers defining the many-body states. To illustrate this point, we consider only vertical optical transitions between the valence and conduction bands, described by the Hamiltonian term given in Eq.~\eqref{ham3}, and neglect Coulomb many-body effects. The quantum-state space then splits into noninteracting subspaces for each wave vector, with only states having the same wave vector but different band indices being coupled. In this case, the JWT yields the multi-qubit Pauli strings, see Eqs.~\eqref{jwt1}--\eqref{jwt2}, and the time evolution generated by the corresponding operators requires exponentiation of these multi-qubit Pauli strings. The exponentiation of a chain of $Z$ operators, which applies a phase shift conditioned on the parity of the corresponding qubits, can be implemented using a sequence of CNOT gates (a CNOT ladder) followed by an appropriate phase rotation. The CNOT sequence is then reversed to uncompute the parity. As shown in Fig.~\ref{fig:orderings}, the number of CNOT gates depends on the difference between the initial and final states associated with the operator $a_j^{\dagger}a_i$ and, consequently, on the ordering of the quantum numbers. If the ordering corresponds to our convention defined by Eq.~\ref{wf}, the number of CNOT gates is $2K$ per time step, that is linearly dependent on $K$ (see Fig.~\ref{fig:orderings}a). For the case shown in Fig.~\ref{fig:orderings}b, where the quantum states are grouped according to their band indices, the number of CNOT gates scales as $K(K-1)$,  i.e. quadratically with $K$. The inclusion of Coulomb effects can further increase the number of two-qubit gates; however, this increase remains polynomial in the number of qubits. Since the light–matter interaction Hamiltonian in Eq.~\eqref{ham3} contains a pair of hopping operators, these estimates need to be multiplied by a factor of two, thus $\alpha=4$.

The overall number of local gates per time step, $\beta$, is determined by the kinetic-energy operator, Eq.~\ref{ham1}, together with the local rotations associated with interaction terms, such as the $R_z$ gates shown in Fig.~\ref{fig:orderings}. The total number of local rotational gates $U$ per time step and per wave vector is 19 plus 2 gates with random parameters used to model dephasing. The gate count per time step and per wave vector with light-matter interaction and without Coulomb coupling is summarized in Table~\ref{tab:count}.

In this work, the simulations were performed using 50 $k$-points and 700 time steps. For the non-interacting case (i.e., without Coulomb effects), this corresponds to 100 qubits and a maximum number of gates of $N_{\mathrm{gates}} = 875050$. The approximate circuit depth can be estimated as $N_{\mathrm{gates}}/(2K) = 8750.5$. The actual depth reported by the compiler is 11201, with 16 gate increments per time step per qubit. The deviation between our estimate and the actual circuit depth arises from the assumption of an equal distribution of gates among the qubits, which is not the case in practice. In particular, there is an imbalance in the number of gates applied to qubits initialized in occupied (valence-band) and unoccupied (conduction-band) states. 

The estimated resource requirements are comparable to the capabilities of the IBM Eagle quantum processor. Note that these resource estimates are obtained from the quantum compiler with the optimization level set to zero. In practice, modern quantum compilers can significantly reduce the number of gates required for a given backend through compiler-level optimizations. Such optimizations can, for example, eliminate pairs of gates that cancel each other and optimize qubit routing, potentially reducing the scaling of the required number of two-qubit gates from polynomial to linear with the number of qubits, as illustrated in Fig.~\ref{fig:orderings}.

The quantum simulations were performed with and without realistic noise models. The realistic noise models are based on a combination of different noise types, with parameters obtained from the snapshots of the IBM Quantum Eagle processor (ibm-kyoto backend) to provide an illustration of the performance expected from a NISQ era device.

The classical simulations were performed by solving the semiconductor Bloch equations using a fourth-order Runge-Kutta method.

Quantum simulation of semiconductor spectroscopy is highly parallelizable, with distinct levels of parallelization determined by the QPU architecture.

First, as shown in Fig.~\ref{fig:steps}, the evaluation of observable expectation values at each time point is independent and can therefore be performed in parallel. However, subsequent steps require an increase in computational time, so the steps are not equivalent in terms of computational cost.

A second level of parallelization arises from the statistical-mechanical structure of the problem. Finite-temperature effects are incorporated by sampling the initial distribution and reconstructing the density matrix $\rho_S$, representing a statistical ensemble of initial states. Each ensemble element can then be simulated independently, with final results obtained by weighted averaging.

Finally, the most efficient parallelization is achieved over wave vectors. This is possible when each $\mathbf{k}$ mode evolves independently under a Liouville--von Neumann equation, as in the single-particle approximation. When many-body effects are included, different $\mathbf{k}$ modes become coupled, which in the quantum algorithm manifests as entanglement between modes. In this regime, efficient simulation requires QPU architectures capable of handling entangled parallel subspaces.

The results of all computational experiments reported in this work can be reproduced using the code available at \href{https://github.com/freude/sbe_qc}{https://github.com/freude/sbe\_qc}.

\begin{table}[t]
\caption{Gate count for the quantum circuit simulating the time propagation of a semiconductor system with light--matter interaction in the semiclassical approximation.}
\label{tab:count}
\begin{tabular}{lc}
Quantity & Value \\
\hline
$\alpha$ (CX-gates per $k$ per time step) & 4 \\
$\beta$ (U-gates per $k$ per time step) & 19 \\
$\nu$ (randomized gates per $k$ per time step) & 2 \\
$n_0$ (gates for initialization per $k$) & 1 \\
Circuit depth increment per time step & $16$ \\
\hline
\end{tabular}
\end{table}

\begin{figure}[t]
    \centering
    \includegraphics[width=\linewidth]{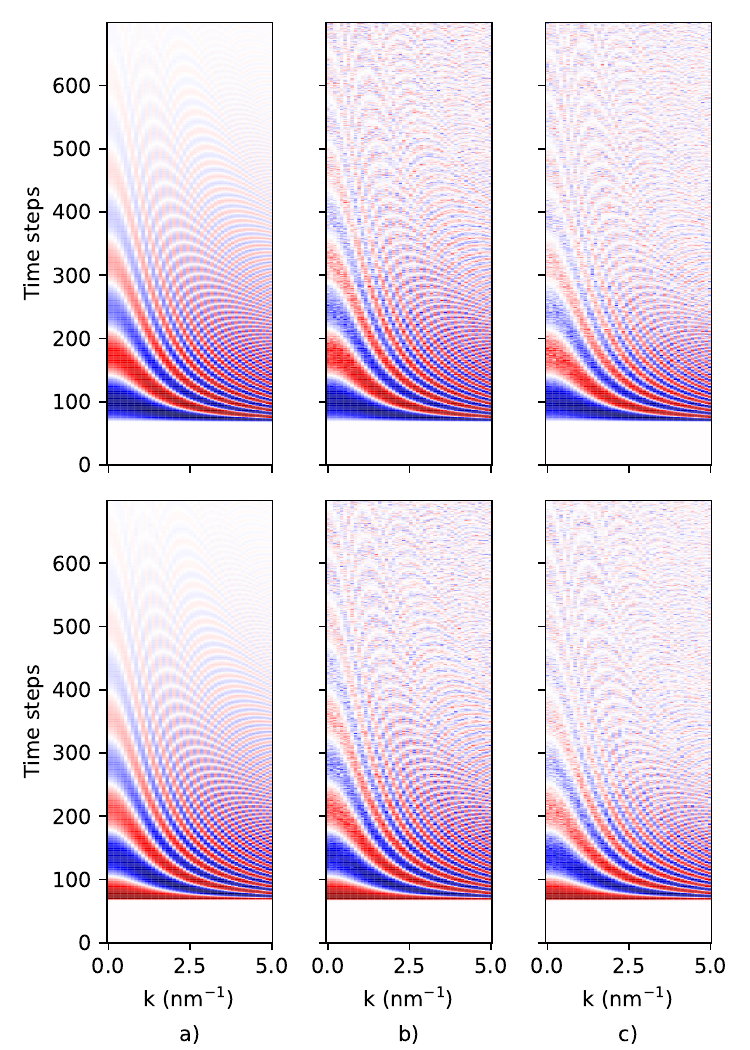}
    \caption{Real (upper row) and imaginary (bottom row) parts of the microscopic polarization in semiconductor medium induced by an ultra-short Gaussian optical pulse. The results are presented for a) classical simulations, b) noiseless quantum simulations, and c) quantum simulations with realistic noise models.}
    \label{fig:dynamics}
\end{figure}

\section{Results: use case of G\MakeLowercase{a}A\MakeLowercase{s} linear absorption spectra}

\subsection{Linear response, $T=0$ K}

The results of the time-domain simulation of the induced dipole moment in a semiconductor interacting with EM field at $T=0$ K are shown in Fig.~\ref{fig:dynamics} for three simulation approaches: classical simulations, noiseless quantum simulation and quantum simulations with realistic noise models. The horizontal axis in these plots indicates discretized values of the wave-vector magnitude. Each wave vector requires two qubits for the quantum simulation. Results are shown for $K=60$ points (120 qubits), compatible with existing IBM Quantum Eagle processors.

The simulation domain consists of 700 Trotter steps and 50 wave vectors. The phenomenological dephasing energy is set to 25 meV for all cases. To keep the error associated with the Lie–Trotter product formula reasonably small, the unitary rotation of the quantum states should remain sufficiently small within a step. At the same time, the number of steps must be sufficient to capture the dynamical evolution of the system, since the size of the time domain determines the spectral resolution in the frequency domain. In addition, sufficiently large time domains ensure that the microscopic polarization decays back to its initial value as a result of the dephasing process, thereby avoiding Gibbs oscillations in the spectral characteristics and enabling zero-padding in the Fourier transform.

A key feature of quantum time-dependent simulations is the ability to map ultrafast dynamics onto timescales that are suitable for qubit manipulation and compatible with qubit coherence times through the use of an appropriate scaling factor. For this reason, it is more meaningful to measure time in terms of Trotter steps rather than absolute time units, since the actual program execution time depends on the specific quantum computing platform. For example, consider an electron-field oscillation with an energy of 0.3 eV. The corresponding oscillation period is 13.8 fs. Suppose we aim to simulate a total time interval of 65 fs, which corresponds to approximately five oscillation periods. Using 700 Trotter steps for this interval yields a time increment of 0.093 fs per step, or about 148 steps per period. This scaling can then be used to express the probe pulse parameters in units of Trotter steps. The results presented in Fig.~\ref{fig:dynamics} have been obtained using an ultrashort Gaussian electromagnetic pulse of the form:
\begin{equation}
    E(t) = E_0 \exp\!\left[-\frac{(t - t_0)^2}{2\sigma^2}\right].
\end{equation}
The parameters are chosen as $E_0 = 2 \times 10^8~\mathrm{V/m}$, $t_0 = 70$ Trotter steps (corresponding to 6.51 fs), and $\sigma = 3.5$ Trotter steps (0.3255 fs). In experimental femtosecond spectroscopy, the value of $\sigma$ is typically one order of magnitude larger. However, in quantum simulations, we can afford to employ shorter pulses, which broaden the spectral bandwidth and enable the excitation of electrons with higher energies. The amplitude is chosen to ensure that the qubit system operates in the linear optical regime, where the populations and corresponding qubit states are not fully flipped by the pulse. Instead, the pulse induces only small perturbations, generating entanglement between qubits and introducing a slight phase shift relative to the initial state.

Qualitatively, the quantum simulations show good agreement with the classical simulation results. The oscillation frequency depends on the wave vector and is determined by the transition energy between the conduction and valence bands defined by the band structure. The oscillation amplitude decays due to dephasing processes induced by scattering events in the semiconductor. The small discrepancies between the classical and noiseless quantum simulations is attributed to quantum fluctuations arising from a finite number of measurements (shots), reflecting the probabilistic nature of quantum measurement outcomes. These errors can be reduced by increasing the number of shots. The inclusion of realistic quantum noise characteristic of NISQ-era quantum computers leads to extra fluctuations in the simulation output.

\begin{figure}[t]
    \centering
    \includegraphics[width=\linewidth]{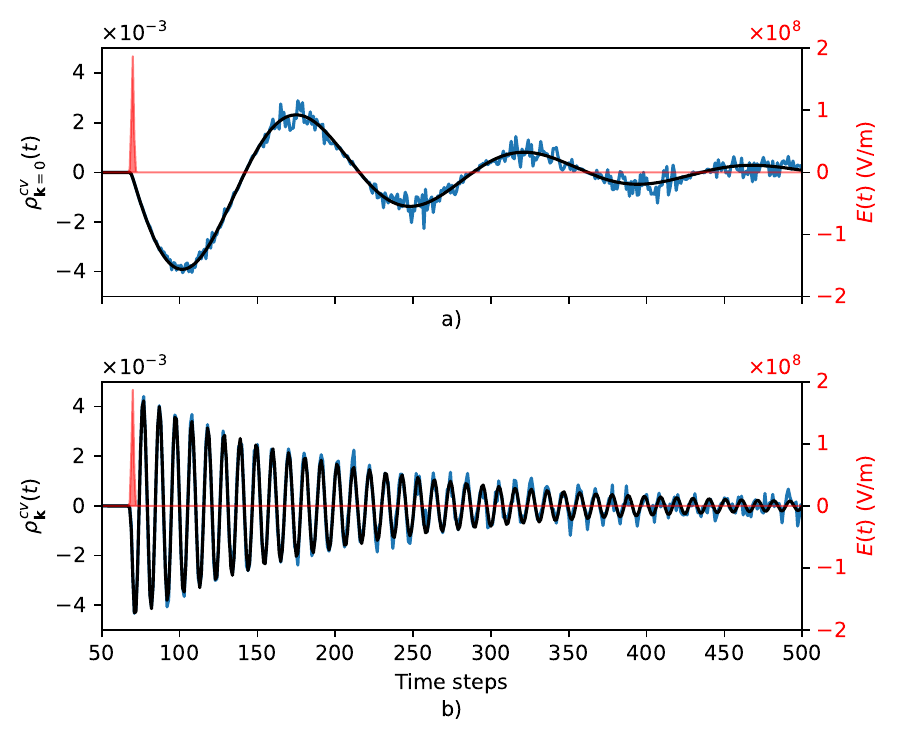}
\caption{Comparison of classical (black line) and quantum (blue line) simulations of the time evolution of the microscopic polarization for two wave vectors: (a) $k = 0\,\mathrm{nm}^{-1}$ and (b) $k = 5\,\mathrm{nm}^{-1}$ in a noiseless environment using $10^4$ shots. The red trace indicates the temporal profile of the probe pulse.}
    \label{fig:1d_no_noise}
\end{figure}

\begin{figure}[t]
    \centering
    \includegraphics[width=\linewidth]{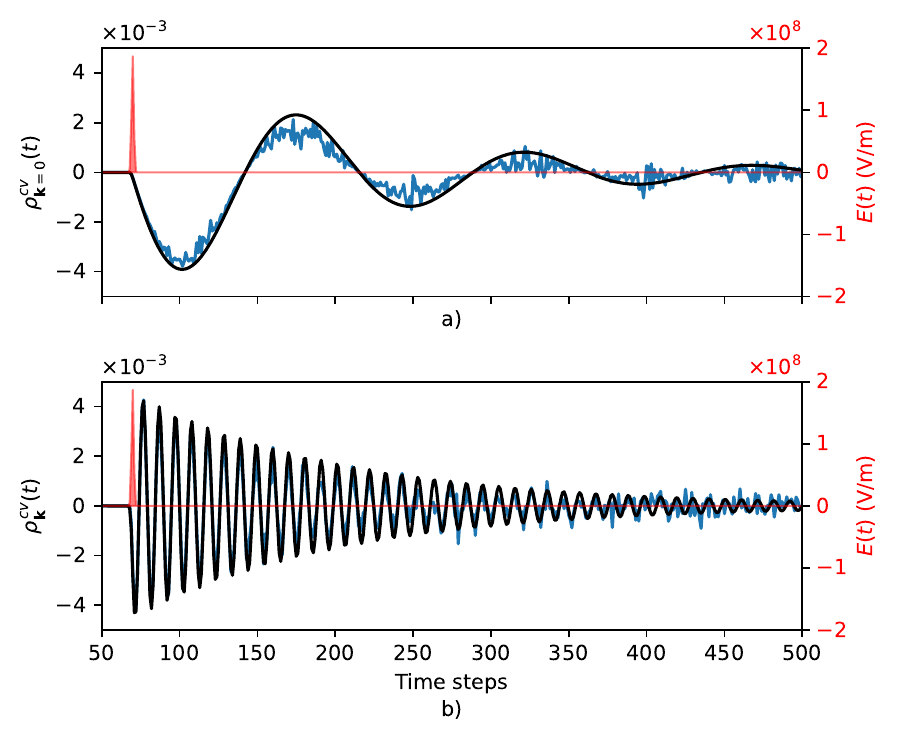}
    \caption{Comparison of classical (black line) and quantum (blue line) simulations of the time evolution of the microscopic polarization for two wave vectors: (a) $k = 0\,\mathrm{nm}^{-1}$ and (b) $k = 5\,\mathrm{nm}^{-1}$, incorporating a realistic noise model using $10^4$ shots. The red trace indicates the temporal profile of the probe pulse.}
    \label{fig:1d}
\end{figure}

\begin{figure}[t!]
    \centering
    \includegraphics[width=\linewidth]{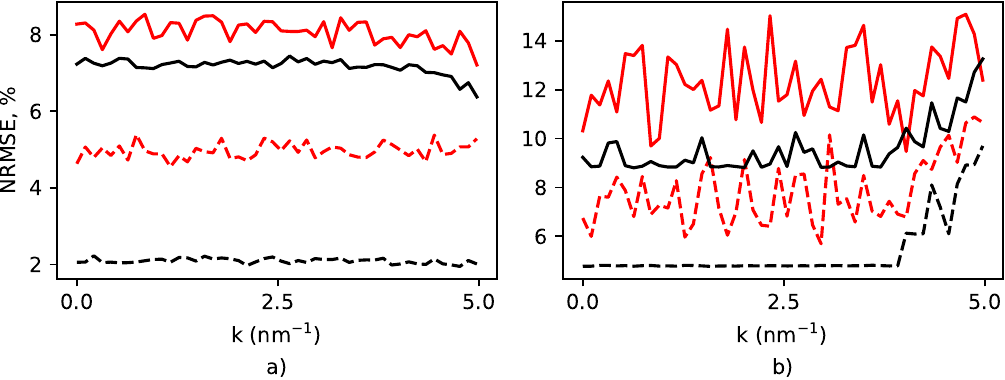}
    \caption{Wave-vector dependence of the normalized root-mean-square error for the quantum simulations of microscopic polarization: (a) absolute value and (b) phase. Results are shown for $10^3$ (red lines) and $10^4$ (black lines) shots for noiseless simulations (dashed lines) and simulations with realistic noise (solid lines).}
    \label{fig:error}
\end{figure}

\begin{figure*}[t!]
    \centering
    \includegraphics[width=0.8\linewidth]{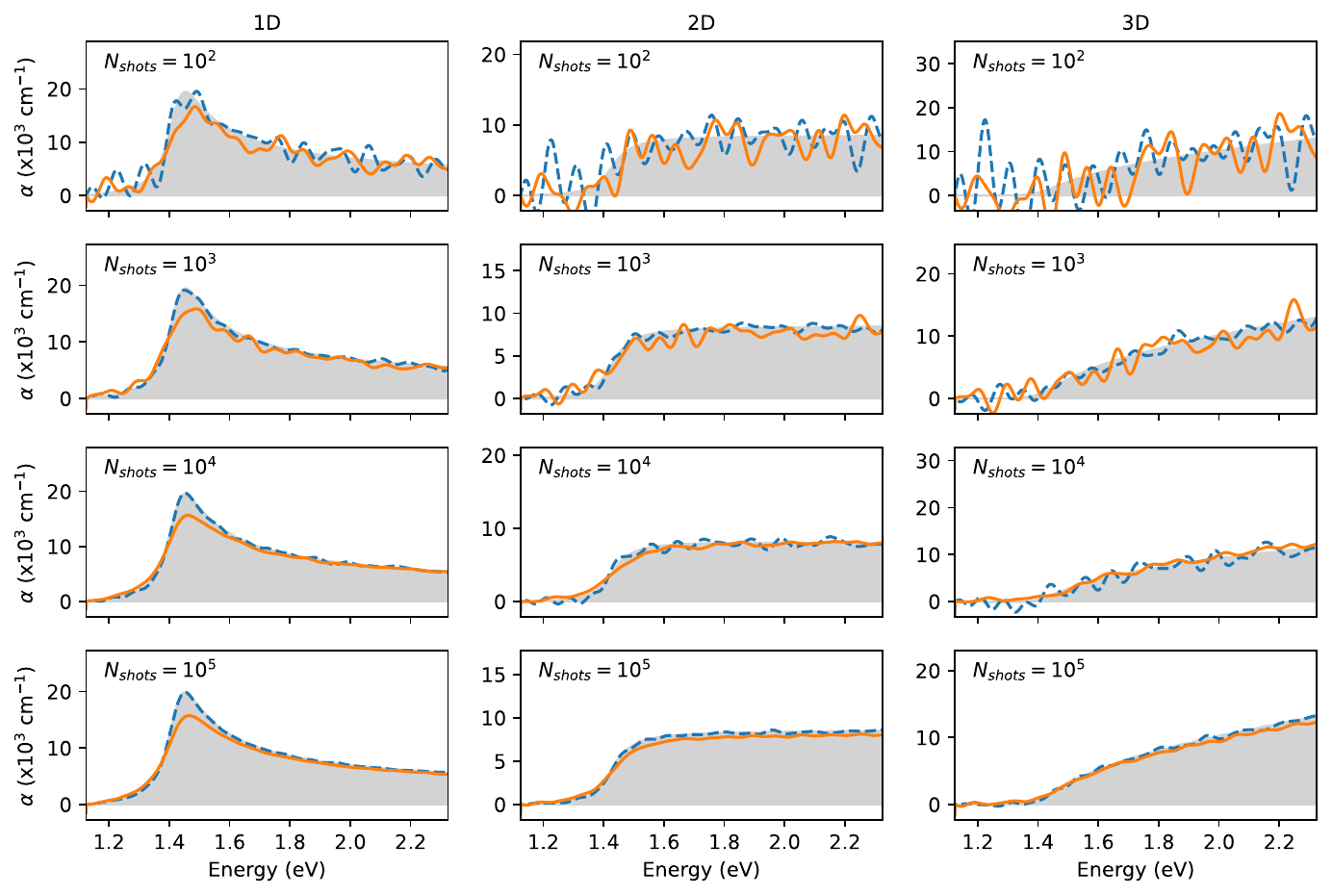}
    \caption{Linear absorption spectra for GaAs obtained using classical simulations (shaded area), noiseless quantum simulations (blue dashed line) and quantum simulations with realistic noise (orange solid line) for 1D, 2D and 3D semiconductor media. Each row of panels from top to bottom shows results obtained with an increasing number of shots for the execution of the quantum algorithm.}
    \label{fig:abs_spectra}
\end{figure*}

By examining the time evolution for several wave vectors (see Fig.~\ref{fig:1d_no_noise} and Fig.~\ref{fig:1d}), one can clearly distinguish between fluctuations due to finite sampling and those arising from quantum noise. In the former case, fluctuations occur around a mean value that exactly matches the classical simulation. In the latter case, quantum noise alters the mean value, leading to deviations in both the amplitude and phase of the oscillations. We now quantify the excess error caused by modelled hardware noise relative to the sampling error alone.  To this end, we evaluate the time-domain normalized root mean square error (NRMSE) of the complex-valued microscopic polarization between the quantum and classical simulations for different wave vectors and two different numbers of shots, as shown in Fig.~\ref{fig:error}.  The error is normalized by the maximum peak-to-peak amplitude. As expected, NRMSE decreases with number of shots, and the inclusion of realistic noise leads to larger NRMSE values for both the absolute value and phase of the microscopic polarization for the same number of shots. For the absolute value,  NRMSE is relatively uniform across the wave vectors. In contrast, the phase exhibits a stronger dependence on the wave vector, with the noise-induced error generally increasing at larger k. This behavior can be attributed to the higher transition frequencies associated with larger wave vectors, for which the quantum state undergoes more substantial changes during each Trotter step, making the phase more susceptible to noise. 

For $10^3$ shots, the mean difference in NRMSE between the noisy and noiseless simulations is $3.11\%$ for the absolute value of the microscopic polarization and $4.50\%$ for its phase. For $10^4$ shots, this difference increases for the absolute value, reaching $5.09\%$, while remaining almost unchanged for the phase, at $4.19\%$. This behavior indicates that a substantial component of the noise-induced error in the absolute value does not vanish with increasing the number of samples. This can be attributed to the fact that hardware noise modifies the mean value of the measured quantity.

The electric susceptibility and linear absorption spectra are obtained from the microscopic polarization by integrating over the wave vector (see Eq.\eqref{polarization}) and subsequently applying the corresponding Fourier transform, as prescribed by Eqs.\eqref{susceptibility}--\eqref{absorption}. The resulting spectra, obtained within the axial approximation for different dimensionalities, are shown in Fig.~\ref{fig:abs_spectra}. In this approximation, the semiconductor medium is assumed to be isotropic. On the absorption spectra, hardware noise manifests itself as an additional broadening. Moreover, with increasing dimensionality, we observe more fluctuations. This is because we use the axial approximation and the formula defined by Eq.~\eqref{axial_approximation}. This formula contains a factor of $k^{D-1}$, where $k$ is the wave vector modulus and $D$ is the dimensionality. This enhances the contribution from large wave vectors, for which the dynamics exhibit rapid oscillations and are therefore more strongly affected by hardware noise.

\subsection{Quasi-equilibrium, finite temperature}

All results presented above were computed assuming thermal equilibrium at 
$T = 0$ K. We next test quantum simulations for non-equilibrium dynamics and finite temperatures. For this, we use the quasi-equilibrium approximation, assigning distinct Fermi levels for electrons in conduction and valence bands, assuming that electrons and holes each remain in their own thermodynamic equilibrium and do not exchange energy. This approximation can describe population inversion, which is a necessary requirement for the stationary generation of semiconductor light-emitting diodes and lasers \cite{chow1999semiconductor, piprek2005optoelectronic}. A measure of the population inversion can be the Fermi wave vector. For direct-band-gap semiconductors at $T=0$~K, the Fermi wave vector is zero when the electron and hole Fermi levels lie within the band gap, and it begins to increase when the Fermi levels cross the band edges and enter the energy bands.

\begin{figure}[t!]
    \centering
    \includegraphics[width=\linewidth]{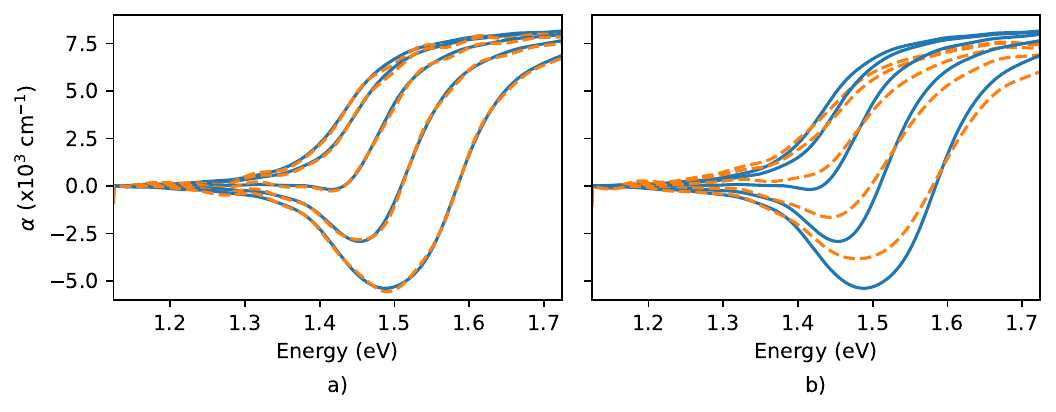}
    \caption{Gain spectra of 2D GaAs computed using classical simulations (blue solid line) and quantum simulations (orange dashed line) for different values of the Fermi wave vector. Results are shown for (a) noiseless simulations and (b) simulations including realistic noise modeled from snapshots of NISQ-era quantum devices. The gain increases with the Fermi wave vector $k_f=0.0, \; 0.5, \; 1.0, \; 1.5, \; 2.0$ nm$^{-1}$.}
    \label{fig:gain_spectra}
\end{figure}

\begin{figure}[t!]
    \centering
    \includegraphics[width=\linewidth]{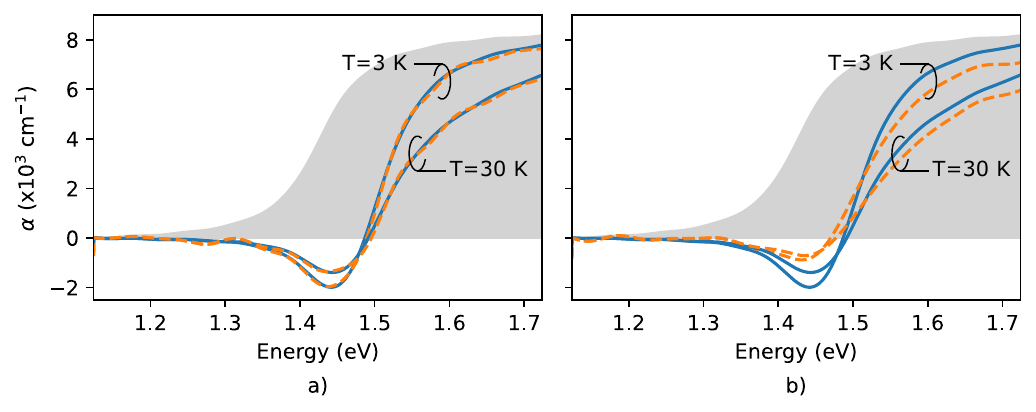}
    \caption{Gain spectra of 2D GaAs computed using classical simulations (blue solid line) and quantum simulations (orange dashed line) for two different temperatures $T=3$ and $30$ K and population inversion $k_f=1$ nm$^{-1}$.  Results are shown for (a) noiseless simulations and (b) simulations including realistic noise modeled from snapshots of NISQ-era quantum devices. For reference, the shaded area shows the absorption spectrum obtained for the system in its ground state at $T=0$~K and $k_f=0$ nm$^{-1}$.}
    \label{fig:gain_spectra_tempr}
\end{figure}

The results presented in Fig.~\ref{fig:gain_spectra} illustrate the emergence of negative absorption (optical gain) in 2D GaAs at $T=3$~K as the population inversion increases. Also, to highlight the finite-temperature effects in the simulations, we model the reduction of the peak gain resulting from an increase in temperature from 3 to 30 K (see Fig.~\ref{fig:gain_spectra_tempr}) while keeping the population inversion constant. The increase in temperature leads to a redistribution of charge carriers within the energy bands.

In both cases, our quantum simulation methodology shows good quantitative agreement with classical simulations (see Fig.~\ref{fig:gain_spectra}a and Fig.~\ref{fig:gain_spectra_tempr}a) when performed on a noiseless quantum computer, provided that a sufficiently large number of shots is used ($10^5$ for the results presented here). However, when realistic noise characteristic of NISQ-era quantum computers is taken into account, the agreement deteriorates (see Fig.~\ref{fig:gain_spectra}b and Fig.~\ref{fig:gain_spectra_tempr}b): the simulated spectra exhibit additional broadening of the gain spectra and a reduced peak gain.

\section{Perspectives for quantum advantage}

\subsection{Expected quantum advantage and hierarchy problem}

In the proposed quantum simulation framework, the two-band semiconductor model requires $2K$ qubits, where  $K$ denotes the number of wave vectors in the discretized Brillouin zone. If the number of Trotter steps is $N$, the error scales as $\mathcal{O}\!\left( \frac{t^2}{N} \right)$ with number of steps, where $t$ is the time interval of interest. Therefore, increasing the number of steps systematically enhances the accuracy of the approximation. The number of single-qubit gates in this case scales as $\mathcal{O}\!\left(N^2 K \right)$.  In the case of pair-wise Coulomb interactions, the number of two-qubit gates scales as $\mathcal{O}\!\left(N^2 K^2 \right)$ without any approximations. The $N^2$ scaling arises because the quantum state is not reused between destructive measurements at each time step, and after such measurements the initial quantum state must be prepared again and propagated up to the previous measurement point. In the future, if qubit resources are not a limiting factor, we anticipate that a more practical approach will be to reuse quantum information from each time step by leveraging the phase kickback effect and/or encoding the entire time trajectory into a single quantum state, known as a \emph{history state}, similarly to the approaches proposed for quantum algorithms for ordinary differential equations \cite{Berry2017}. The quantum trajectory method used to simulate non-unitary dissipative and scattering-assisted dynamics requires multiple executions of circuits with randomly sampled gates, leading to additional sampling overhead that may be interpreted as an increased shot count. This overhead can be reduced through the use of so-called dynamic circuits \cite{11425010}, which replace repeated circuit executions with deeper circuits incorporating intermediate measurements and feedforward operations, thereby lowering the number of samples required by the randomized algorithm.

To compare the computational complexity with that of classical simulations, we note that the density matrix in Eq.~\eqref{eq:liouville} is of dimension $2^{2K} \times 2^{2K}$ for two energy bands. Consequently, the dimension of the Liouville space is $2^{4K}$. Therefore, the Liouville--von Neumann equation constitutes a system of $2^{4K}$ coupled ordinary differential equations. When integrated numerically (e.g., using the Euler method), the computational complexity scales as $\mathcal{O}\!\left( N\,2^{4K} \right)$ for uncoupled or weakly coupled equations, and as $\mathcal{O}\!\left( N\,2^{8K} \right)$ for a strongly coupled system, where $N$ is the number of time steps. 

At first glance, this analysis suggests that quantum simulations offer an exponential advantage for sufficiently dense k-point grids, while exhibiting slightly worse scaling in the number of time steps (polynomial versus linear) in the current implementation. However, this estimate is undoubtedly overstated, as classical simulations in practice do not typically involve propagating the entire many-body density matrix in time. Instead, they generally focus on the time evolution of the single-particle density matrix needed to compute the observables of interest. Such time evolution is, in turn, affected by many-particle correlations, giving rise to the so-called hierarchy problem. In this, the efficiency of classical simulations can be improved by introducing various types of approximations, at the expense of accuracy. In the single-particle approximation, the semiconductor Bloch equations comprise $3K$ independent ordinary differential equations, thus teh computational complexity is $\mathcal{O}\!\left( N K \right)$

The hierarchy problem in quantum kinetics describes the situation when many-body interactions cause the equation of motion for the single-particle density matrix to depend on two-particle correlations, whose dynamics, in turn, depend on three-particle correlations, and so on, resulting in an infinite hierarchy of coupled equations:

\begin{equation}
i \hbar \frac{d \langle N \rangle}{dt} = T \left[ \langle N \rangle \right] +  V \left[ \langle N + 1 \rangle \right],
\end{equation}
where $\langle N \rangle$ is $N$-particle correlation function, which, for a fermionic system with conserved particle number, is given by the expectation value of a product of $N$ creation and $N$ annihilation operators, $T$ and $V$ are known functionals.

In classical simulations, this problem is addressed by applying a cluster expansion to products of creation and annihilation operators and truncating the hierarchy at a desired level of accuracy (see \cite{KIRA2006155}  for more details). For instance, the cluster expansion of a quantity involving four operators,  $\langle N= 4 \rangle$, reads:

\begin{equation}
\begin{aligned}
\langle  a_A^{\dagger} a_B^{\dagger} a_C a_D \rangle &= \langle  a_A^{\dagger} a_D \rangle  \langle  a_B^{\dagger} a_C  \rangle \\
& -\langle  a_A^{\dagger} a_C \rangle  \langle  a_B^{\dagger} a_D  \rangle \\ 
& + \Delta \langle  a_A^{\dagger} a_B^{\dagger} a_C a_D \rangle 
\end{aligned}
\label{cluster}
\end{equation}

This equation shows that the four-operator correlation function can be expressed as a sum of all possible products of two-operator (single-particle) correlation functions, plus a remainder term denoted by a capital delta. Neglecting this remainder yields a closed system of semiconductor Bloch equations within the Hartree–Fock approximation. In the Hartree--Fock limit, the number of equations remains $3K$, although they become coupled as a result of long-range Coulomb interactions. As a result, the computational complexity for solving the kinetic equations numerically scales as $\mathcal{O}\left(N K^2\right)$. Going beyond the Hartree–Fock approximation generally requires solving the kinetic equations for four-operator correlation functions. These equations involve products of six operators on the right-hand side. Applying the cluster expansion decomposes these six-operator products into combinations of two- and four-operator correlation functions, together with a remainder term that can be neglected under an appropriate approximation. In this case, the number of kinetic equations, the number of terms on their right-hand sides, and the coupling between them all increase, thereby raising the computational complexity of the algorithm to at least $\mathcal{O}\left(N K^4\right)$.   Thus, by going higher up the hierarchy of electron correlations, we improve the accuracy, but with a significantly higher computational complexity at each successive level.

In contrast, the quantum-computing framework presented in this work enables the time propagation of the fully correlated many-body quantum state, thereby avoiding the hierarchy problem. Our analysis shows that an advantage in computational scaling with respect to the number of wave vectors is achieved beyond the Hartree–Fock approximation.

\subsection{Relation to classical coupled oscillators and synchronization}

A simple way to demonstrate the advantage of quantum simulations for many-body problems in semiconductor spectroscopy is to map the problem at hand onto a system of coupled classical oscillators, for which a quantum advantage has been formally established in Ref.~\cite{PhysRevX.13.041041}. Indeed, in the linear case, the equations of motion for the microscopic polarization  in the Hartree-Fock approximation becomes equivalent to the equation of motion of a system of $K$ coupled, driven, damped classical oscillators. 

Note that, within the Hartree-Fock approximation, the Rabi frequency in Eq.~\eqref{HF_Rabi} contains a term that couples different wave vectors and, hence, oscillators. This can be easily checked by the substitution in Eq.~\eqref{sbe1}:
\begin{equation*}
    p_{\mathbf{k}} = \sqrt{m_{\mathbf{k}} \omega_{\mathbf{k}}}Q_{\mathbf{k}} - \frac{i}{\sqrt{m_{\mathbf{k}} \omega_k}}P_{\mathbf{k}}
\end{equation*}
where $\mathbf{k}$ now denotes the oscillator index, $Q_{\mathbf{k}}$ and $P_{\mathbf{k}}$ are generalized coordinates and momenta, respectively, $m_{\mathbf{k}}$ is the mass of the classical particle, and $\omega_{\mathbf{k}} =(\varepsilon_{c,\mathbf{k}} - \varepsilon_{v,\mathbf{k}})/\hbar$ is the oscillation frequency of $\mathbf{k}$-th oscillator  expressed in terms of the electron and hole energies. If the masses of the classical particles are chosen to be $m_{\mathbf{k}} = 1 / \omega_{\mathbf{k}}$, this expression becomes simpler:
\begin{equation}
    p_{\mathbf{k}} = Q_{\mathbf{k}} - i P_{\mathbf{k}}
    \label{sub}
\end{equation}
leading to the set of classical equations for coupled oscillating particles:
\begin{equation}
\frac{d Q_{\mathbf{k}} }{dt}  =  \omega_{\mathbf{k}} P_{\mathbf{k}} + \frac{1}{\hbar} \sum_{\mathbf{q} \neq \mathbf{k}} V_{|\mathbf{k} - \mathbf{q}|}  P_{\mathbf{q}} - \gamma Q_{\mathbf{k}}, 
    \label{hams1}
\end{equation}
\begin{equation}
\frac{d P_{\mathbf{k}} }{dt}  =  -\omega_{\mathbf{k}} Q_{\mathbf{k}} - F(t) - \frac{1}{\hbar}  \sum_{\mathbf{q} \neq \mathbf{k}} V_{|\mathbf{k} - \mathbf{q}|}  Q_{\mathbf{q}}  -  \gamma P_{\mathbf{k}}, 
    \label{hams1}
\end{equation}

These equations of motion correspond to Bateman’s dual Hamiltonian for dissipative systems  \cite{DEKKER19811}, providing a workaround for describing non-Hamiltonian dynamics within a Hamiltonian framework. Without interactions this recovers Newton's equation for uncoupled oscillators driven by the force proportional to E(t):
\begin{equation}
\frac{1}{\omega_{\mathbf{k}}} \frac{d^2 Q_{\mathbf{k}} }{dt^2} + \frac{2 \gamma}{\omega_{\mathbf{k}}}  \frac{d Q_{\mathbf{k}} }{dt} + \frac{\left(\omega_{\mathbf{k}}^2 + \gamma^2 \right)}{\omega_{\mathbf{k}}} Q_{\mathbf{k}} =  E\left( t \right) .
    \label{newton}
\end{equation}
The second term on the right-hand side of Eq.~\eqref{newton} describes the damping of the oscillations, with a friction coefficient $2\gamma/\omega_{\mathbf{k}}$, while the term $E(t)$ on the right-hand side can be interpreted as a time-dependent driving force.

This maps the problem of quantum simulation of linear semiconductor spectroscopy in the Hartree-Fock approximation to that of quantum simulation of coupled classical oscillators. For the latter problem, an exponential speed up has been theoretically demonstrated in Ref.~\cite{PhysRevX.13.041041}. In Ref.~\cite{PhysRevX.13.041041}, the values of $Q_{\mathbf{k}}$ and $P_{\mathbf{k}}$ are encoded in the qubit populations, and, similarly to the present work, simulating $K$ oscillators requires $2K$ qubits.

For classical oscillators, the coupling manifests itself as synchronization phenomena, and Eq.~\eqref{sbe1} with coupling takes the form of the Kuramoto model \cite{RevModPhys.77.137}. The excitonic peaks in the absorption spectrum, arising from Coulomb coupling, correspond to collective oscillation modes of the ensemble of oscillators. Note that potential advantages of quantum simulation of the Kuramoto model have been recently discussed in Ref.~\cite{leditto2026} in the context of network theory.

\subsection{Application to nonlinear optical response}

Unlike conventional classical simulations, where the linear-response regime is enforced by neglecting carrier population dynamics and assuming time-independent distribution functions, the proposed quantum-simulation framework is intrinsically nonlinear. Population dynamics are naturally included, and no fundamental restriction is imposed on the driving-field amplitude. In this work, the linear regime is recovered by limiting the electromagnetic-field strength. This intrinsic nonlinearity, combined with the ability to perform measurements for individual wave vectors along selected directions, makes the approach promising for studies of nonlinear and ultrafast optical phenomena.

Another possible direct extension is the full quantization of the electromagnetic field, enabling the simulation of genuine quantum light--matter dynamics relevant to cavity quantum electrodynamics and quantum-optical systems.

\section{Conclusions}

We have demonstrated that quantum simulations accurately reproduce key features of semiconductor spectroscopy, including temperature-dependent spectral  characteristics, optical gain, dephasing, and spectral broadening. In the noiseless limit, quantum and classical calculations of absorption and gain spectra are in excellent agreement. With realistic NISQ noise, the simulations remain robust, exhibiting additional spectral broadening and reduced gain while preserving qualitative features. 

Although demonstrated here in the single-particle regime, where no quantum advantage is expected, the framework can be directly extended to interacting many-body and strongly non-equilibrium systems. We have demonstrated that the linear optical response of semiconductors within the Hartree–Fock approximation is equivalent to the dynamics of a system of coupled classical oscillators, for which exponential speed up has been formally demonstrated under certain assumptions in~\cite{PhysRevX.13.041041}. 

Going beyond the Hartree-Fock approximation offers an even greater quantum advantage, as the algorithmic complexity of the quantum simulation remains unchanged, whereas classical simulations face the quantum kinetics hierarchy problem.  Simulations that include many-body effects are expected to be more sensitive to noise, as they require a larger number of two-qubit gates. In this case, quantum error mitigation techniques are required.

We have considered two distinct sources of dephasing: the first is intrinsic to the physical systems being simulated and is intentionally parameterized and introduced into the model in a controllable manner; the second is parasitic dephasing introduced by the quantum hardware, which is unavoidable and cannot be directly controlled without applying error-mitigation techniques. Without error mitigation, the modeled hardware noise manifests as additional broadening of the simulated optical characteristics.

Beyond its application to spectroscopy, owing to its systematic scalability with the number of wave vectors, energy bands, and interaction channels, the proposed framework can also serve as a flexible benchmark for quantum hardware. The availability of experimentally measurable reference spectra further supports semiconductor spectroscopy as a promising platform for standardized validation of quantum computing systems.

These results position semiconductor spectroscopy as a compelling and physically relevant application domain for quantum simulation of many-body optical phenomena in condensed matter.

\bibliographystyle{quantum}
\bibliography{main}

\end{document}